\documentclass[onecolumn]{IEEEtran}
\usepackage{amsmath,amssymb,bm,cite,color,amsthm,mathrsfs}
\usepackage{enumerate}
\usepackage{hyperref}
\usepackage{cleveref}

\newtheorem{theorem}{Theorem}[section]
\newtheorem{lemma}{Lemma}[section]
\newtheorem{claim}{Claim}[section]

\newtheorem{corollary}{Corollary}[section]
\newtheorem{definition}{Definition}[section]

\newtheorem{example}{Example}[section]
\newtheorem{remark}{Remark}[section]

\newcommand\nc\newcommand
\nc{\cA}{\mathcal{A}}\nc{\cB}{\mathcal{B}}\nc{\cC}{\mathcal{C}}\nc{\cD}{\mathcal{D}}
\nc{\cE}{\mathcal{E}}\nc{\cF}{\mathcal{F}}\nc{\cG}{\mathcal{G}}\nc{\cH}{\mathcal{H}}
\nc{\cI}{\mathcal{I}}\nc{\cJ}{\mathcal{J}}\nc{\cK}{\mathcal{K}}\nc{\cL}{\mathcal{L}}
\nc{\cM}{\mathcal{M}}\nc{\cN}{\mathcal{N}}\nc{\cO}{\mathcal{O}}\nc{\cP}{\mathcal{P}}
\nc{\cQ}{\mathcal{Q}}\nc{\cR}{\mathcal{R}}\nc{\cS}{\mathcal{S}}\nc{\cT}{\mathcal{T}}
\nc{\cU}{\mathcal{U}}\nc{\cV}{\mathcal{V}}\nc{\cW}{\mathcal{W}}\nc{\cX}{\mathcal{X}}
\nc{\cY}{\mathcal{Y}}\nc{\cZ}{\mathcal{Z}}

\nc{\bba}{\mathbf{a}}\nc{\bbb}{\mathbf{b}}\nc{\bbc}{\mathbf{c}}\nc{\bbd}{\mathbf{d}}
\nc{\bbe}{\mathbf{e}}\nc{\bbf}{\mathbf{f}}\nc{\bbg}{\mathbf{g}}\nc{\bbh}{\mathbf{h}}
\nc{\bbi}{\mathbf{i}}\nc{\bbj}{\mathbf{j}}\nc{\bbk}{\mathbf{k}}\nc{\bbl}{\mathbf{l}}
\nc{\bbm}{\mathbf{m}}\nc{\bbn}{\mathbf{n}}\nc{\bbo}{\mathbf{o}}\nc{\bbp}{\mathbf{p}}
\nc{\bbq}{\mathbf{q}}\nc{\bbr}{\mathbf{r}}\nc{\bbs}{\mathbf{s}}\nc{\bbt}{\mathbf{t}}
\nc{\bbu}{\mathbf{u}}\nc{\bbv}{\mathbf{v}}\nc{\bbw}{\mathbf{w}}\nc{\bfx}{\mathbf{x}}
\nc{\bby}{\mathbf{y}}\nc{\bbz}{\mathbf{z}}

\nc{\bbA}{\mathbf{A}}\nc{\bbB}{\mathbf{B}}\nc{\bbC}{\mathbf{C}}\nc{\bbD}{\mathbf{D}}
\nc{\bbE}{\mathbf{E}}\nc{\bbF}{\mathbf{F}}\nc{\bbG}{\mathbf{G}}\nc{\bbH}{\mathbf{H}}
\nc{\bbI}{\mathbf{I}}\nc{\bbJ}{\mathbf{J}}\nc{\bbK}{\mathbf{K}}\nc{\bbL}{\mathbf{L}}
\nc{\bbM}{\mathbf{M}}\nc{\bbN}{\mathbf{N}}\nc{\bbO}{\mathbf{O}}\nc{\bbP}{\mathbf{P}}
\nc{\bbQ}{\mathbf{Q}}\nc{\bbR}{\mathbf{R}}\nc{\bbS}{\mathbf{S}}\nc{\bbT}{\mathbf{T}}
\nc{\bbU}{\mathbf{U}}\nc{\bbV}{\mathbf{V}}\nc{\bbW}{\mathbf{W}}\nc{\bfX}{\mathbf{X}}
\nc{\bbY}{\mathbf{Y}}\nc{\bbZ}{\mathbf{Z}}

\nc{\sA}{\mathsf{A}}\nc{\sB}{\mathsf{B}}\nc{\sC}{\mathsf{C}}\nc{\sD}{\mathsf{D}}
\nc{\sE}{\mathsf{E}}\nc{\sF}{\mathsf{F}}\nc{\sG}{\mathsf{G}}\nc{\sH}{\mathsf{H}}
\nc{\sI}{\mathsf{I}}\nc{\sJ}{\mathsf{J}}\nc{\sK}{\mathsf{K}}\nc{\sL}{\mathsf{L}}
\nc{\sM}{\mathsf{M}}\nc{\sN}{\mathsf{N}}\nc{\sO}{\mathsf{O}}\nc{\sP}{\mathsf{P}}
\nc{\sQ}{\mathsf{Q}}\nc{\sR}{\mathsf{R}}\nc{\sS}{\mathsf{S}}\nc{\sT}{\mathsf{T}}
\nc{\sU}{\mathsf{U}}\nc{\sV}{\mathsf{V}}\nc{\sW}{\mathsf{W}}\nc{\sX}{\mathsf{X}}
\nc{\sY}{\mathsf{Y}}\nc{\sZ}{\mathsf{Z}}

\newcommand{\mathset}[1]{\left\{#1\right\}}

\newcommand{\abs}[1]{\left|#1\right|}
\newcommand{\ceilenv}[1]{\left\lceil #1 \right\rceil}
\newcommand{\floorenv}[1]{\left\lfloor #1 \right\rfloor}
\newcommand{\parenv}[1]{\left( #1 \right)}
\newcommand{\sparenv}[1]{\left[ #1 \right]}

\nc{\set}[1]{\llbracket #1 \rrbracket}

\title{On the Maximum Size of Codes Under the Damerau-Levenshtein Metric}
\author{Zuo~Ye and Gennian~Ge%
\thanks{The research of G. Ge was supported by the National Key Research and Development Program of China under Grant 2020YFA0712100, the National
Natural Science Foundation of China under Grant 12231014, and Beijing Scholars Program.

Z. Ye is with the Institute of Mathematics and Interdisciplinary Sciences, Xidian University, 
Xian 710126, China. Email: yezuo@xidian.edu.cn.

G. Ge is with the School of Mathematical Sciences, Capital Normal University, Beijing 100048, China. Email: gnge@zju.edu.cn.
}
}

\begin{document}
\maketitle

\begin{abstract}
The Damerau–Levenshtein distance between two sequences is the minimum number of operations (deletions, insertions, substitutions, and adjacent transpositions) required to convert one sequence into another. Notwithstanding a long history of this metric, research on error-correcting codes under this distance has remained limited. Recently, motivated by applications in DNA-based storage systems, Gabrys \emph{et al} and Wang \emph{et al} reinvigorated interest in this metric. In their works, some codes correcting both deletions and adjacent transpositions were constructed. However, theoretical upper bounds on code sizes under this metric have not yet been established. This paper seeks to establish upper bounds for code sizes in the Damerau–Levenshtein metric. Our results show that the code correcting one deletion and asymmetric adjacent transpositions proposed by Wang \emph{et al} achieves optimal redundancy up to an additive constant.
\end{abstract}

\section{Introduction}
Let $\bfx=x_1x_2\cdots x_n$ be a sequence over some alphabet $\Sigma$. There are four common operations on sequences: 1) deletion: removing some $x_i$ from $\bfx$; 2) insertion: inserting a symbol from $\Sigma$ into $\bfx$; 3) substitution: replacing some $x_i$ by a symbol in $\Sigma\setminus\{x_i\}$; 4) adjacent transposition (or transposition, for short): swaping neighboring symbols $x_i$ and $x_{i+1}$, where $x_i\ne x_{i+1}$. The Levenshtein distance (or edit distance) between two sequences is the minimum number of insertions, deletions and substitutions needed to transform one sequence into another. The Damerau–Levenshtein distance additionally allows transpositions.

The substitution error and related error-correcting codes have been extensively investigated and well-understood since Shannon's pioneering work \cite{Shannon1948}. The research on insertion/deletion-correcting codes dates back to as early as 1962 \cite{Sellers1962}. In 1966, Levenshtein presented in his seminal work \cite{VL1966} a nearly optimal binary code (VT code) which can correct one insertion/deletion. Building on the code in \cite{VL1966}, Tenengolts in 1984 constructed a non-binary code correcting one insertion/deletion \cite{Tenengolts1984}. Primarily driven by applications in DNA-based storage systems \cite{Yazdi2015IEEE,hec2019}, racetrack memories \cite{YeowMeng2018it,Sima2023it} and document exchange \cite{Cheng2018FOCS,Haeupler2019FOCS}, there has been a significant volume of research focusing on insertion/deletion-correcting codes in the past decade. See \cite{Gabrys2019it,Sima2020it,Sima2021it,Guruswami2021it,Song2023IT,Tuan2024IT,Zuoye202408arXiv} and references therein. Codes simultaneously handling insertions, deletions and substitutions have also been explored \cite{Smagloy2023it,Song2022IT,Sun2024IT,Yuhang202401arXiv}.

To the best of the authors' knowledge, adjacent transposition errors first gained attention because it appeared as one of the four common spelling errors (substitution, insertion, deletion and adjacent transposition) \cite{Damerau1964,Abdel-Ghaffar1998}. This type of errors (known as peak-shifts) also occur in magnetic recording systems \cite{Tahara1976,Wood1989}. There is a sizable body of literature on binary codes correcting transpositions \cite{Hilden1991,Shamai1991it,Ferreira1991it,Kuznetsov1993it,Levenshtein1993it,Mladen2019it,Mladen2024researchgate}.

Although there are a lot of works focusing on codes under the Levenshtein metric or codes combating only transpositions, there are limited results on the interaction between insertions/deletions/substitutions and transpositions. As far as we know, prior to \cite{Ryan2018IT}, the sole existing research pertaining to this subject is \cite{Schulman1999IT}, in which asymptotically good codes correcting insertions, deletions and transpositions were constructed. It is noteworthy that the work \cite{Schulman1999IT} specifically addresses scenarios where the number of errors scales with the code-length.

Prompted by applications in DNA-based storage systems, Gabrys \textit{et al} reinvigorated the investigation of codes capable of combating both deletions and transpositions \cite{Ryan2018IT}. In their work, the number of errors was assumed to be constant compared to the code-length $n$. By combining a variant of the VT code and a code which can correct at most $2\ell$ substitutions, they firstly constructed a binary code correcting one deletion and at most $\ell$ transpositions with $(2\ell+1)\log n+O(1)$ redundant bits. Then they extended the idea and constructed a binary code which can correct one block deletion and one block adjacent transposition. Then lengths of deletion-block and transposition-block were assumed to be the same and at most $b$, which is a constant compared to $n$. The redundancy of the constructed code is $\ceilenv{\log b}\log n+O\parenv{b^2\log\log n}$.

Regarding the binary alphabet $\{0,1\}$, there are two types of transpositions: $0$-right shifts (i.e., $01\rightarrow 10$) and $0$-left shifts (i.e., $10\rightarrow 01$). In \cite{Ryan2018IT}, the authors did not distinguish these two types of transpositions. In some application domains, the two error types may occur with different probabilities \cite{Nunnelley1990}. Following \cite{Ryan2018IT}, Wang \textit{et al} \cite{Shuche2025TCOMM} initiated the study of some variants of Gabrys \textit{et al}'s error model, among which is the one regarding deletions and asymmetric transpositions (that is, assuming different maximum number of $0$-right shifts and $0$-left shifts). They constructed a code with $\parenv{1+t^{+}+t^{-}}\log\parenv{n+t^{+}+t^{-}+1}+1$ bits of redundancy, which can correct one deletion, $t^{+}$ right-shifts and $t^{-}$ left-shifts.

The aforementioned two works predominantly concentrated on constructions of codes. Upper bounds on cardinality of respective codes remain unknown, except the asymptotic upper bound on the maximum size of binary codes correcting $t$ deletions of symbol $0$s and $s$ transpositions given in \cite{Shuche2025TCOMM}. In this paper, we continue this line of research and aim to derive upper bounds on the maximum size of respective codes mentioned above. We first give upper bounds on code sizes when only deletions and transpositions occur. Then we extend the idea to derive an upper bound when all four types of errors may occur. At last, we upper bound the size of codes correcting deletions and asymmetric transpositions. Our results show that the redundancy of a code is at least $t\log n-O(1)$, where $t$, assumed to be a constant, is the total number of different types of errors (insertions, deletions, substitutions and transpositions) that this code can correct. In particular, this confirms that the aforementioned code constructed in \cite{Shuche2025TCOMM} has optimal redundancy up to an additive constant.

This paper is organized as follows. In \Cref{sec_preliminary}, some notations, terminologies and preliminary results are introduced. \Cref{sec_ballsize} focuses on the size of error balls when only deletions and transpositions occur. Based on these results, upper bounds on the size of codes correcting both deletions and transpositions are derived in \Cref{sec_bounddeltrans}. In \Cref{sec_boundblock}, we give an upper bound on the size of codes correcting block deletions and block transpositions. In \Cref{sec_boundextension,sec_boundasy}, we apply part of the idea in \Cref{sec_bounddeltrans} to give upper bounds on the size of codes under the Demerau-Levenshtein metric and of codes correcting deletions and asymmetric transpositions. Finally, \Cref{sec_conclusion} concludes this paper.

\section{Preliminary}\label{sec_preliminary}
For two integers $m$ and $n$ with $m\le n$, define $[m,n]\triangleq\mathset{m,m+1,\ldots,n}$. When $m=1$, we also write $[1,n]$ as $[n]$. For any integer $q\ge 2$, let $\Sigma_q\triangleq\{0,1,\ldots,q-1\}$ denote the $q$-ary alphabet.

Given integers $n\ge 0$ and $q\ge 2$, let $\Sigma_q^{n}$ be the set of all $q$-ary sequences of length $n$. Here $\Sigma_q^0$ consists of the unique empty sequence, which is denoted by $\epsilon$. Let $\Sigma_q^{*}\triangleq\mathop{\cup}\limits_{n\ge0}\Sigma_q^n$, i.e., the set of all sequences of finite length. Sequences in $\Sigma_q^n$ are denoted by bold letters. Given a sequence $\bfx\in\Sigma_q^n$, unless otherwise stated, denote by $x_i$ the $i$-th entry of $\bfx$, for each $1\le i\le n$. The sequence $\bfx$ can therefore be expressed as $\bfx=x_1\cdots x_n$. More generally, for a subset $I=\mathset{i_1,i_2,\ldots,i_k}$ of $[n]$ (where $i_1<i_2<\cdots<i_k$), denote $\bfx_{I}=x_{i_1}x_{i_2}\cdots x_{i_k}$. Clearly, the sequence $\bfx_{I}$ is obtained from $\bfx$ by deleting $n-k$ symbols. We call $\bfx_{I}$ a \emph{subsequence} of $\bfx$. In particular, the subsequence $\bfx_{[i,j]}$ is called a \emph{substring} of $\bfx$, where $1\le i\le j\le n$. For convenience, if $j=i-1$, we regard $\bfx_{[i,j]}$ as the empty sequence. For $0\le s\le n$. Let $\cD_s(\bfx)$ be the set of all subsequences of $\bfx$ of length $n-s$.

The concatenation of two sequences $\bfx,\bby\in\Sigma_q^{*}$ is denoted by $\bfx\bby$.
For two sequences $\bfx,\bby\in\Sigma_q^{n}$, we say that $\bby$ is obtained from $\bfx$ by an \textit{adjacent transposition} (or \textit{transposition} for short) at position $k$ (where $1\le k<n$) if $x_k\ne x_{k+1}$, $y_ky_{k+1}=x_{k+1}x_k$ and $y_l=x_l$ for all $l\ne k,k+1$. In this case, we also denote $\bby=T(\bfx,k)$. Let $n\ge2$ be an integer. For integer $t\ge 0$, denote by $\cT_{t}(\bfx)$ the set of all sequences which can be obtained by applying exactly $t$ transpositions to $\bfx$. Define $\cT_{\le t}(\bfx)\triangleq\cup_{i=0}^t\cT_{i}(\bfx)$, which is the set of all sequences obtained from $\bfx$ by \emph{at most} $t$ transpositions. The set $\cT_{\le t}(\bfx)$ is called the \emph{$t$-transposition ball} centered at $\bfx$. Note that $\bfx\in\cT_{\le t}(\bfx)$ for any $t\ge0$.

In the definition of $\cT_{t}(\bfx)$, the $t$ transpositions can always be posited to occur sequentially. More precisely, for any $\bby\in\cT_{t}(\bfx)$, there must exist $k_1,\ldots,k_t$, such that $\bby=T(\bbz_{t},k_t)$, where $\bbz_i=T(\bbz_{i-1},k_{i})$ for all $i=1,\ldots,t-1$. Here, $\bbz_0=\bfx$. It might be that $\abs{k_i-k_j}>2$ for any $i\ne j$, i.e., all transposed pairs are mutually non-overlapping. In this case, we also say that these $t$ transpositions occur \emph{simultaneously}. In general, a sequence results from $t$ sequential transpositions can not necessarily be obtained by $t$ simultaneous transpositions.
\begin{example}
    Let $\bfx=01021$. Transposing $x_1$ and $x_2$, and $x_4$ and $x_5$ simultaneously, we can obtain the sequence $10012$. If we first transpose $x_4$ and $x_5$ to get sequence $\bby=01012$, and then transpose $y_3$ and $y_4$, we will obtain the sequence $\bbz=01102$. Clearly, $\bbz$ can not result from two simultaneous transpositions.
\end{example}

In this and subsequent two sections, we focus on the interaction between deletions and transpositions. It is convenient to define the following set
$$
\cB_{s,t}(\bfx)=\mathset{\bby\in\Sigma_q^{n-s}:
\begin{array}{c}
     \bby\text{ is obtained from }\bfx\text{ by exactly }s\\
      \text{deletions and at most }t\text{ transpositions}
\end{array}
},
$$
where $\bfx\in\Sigma_q^n$, $t\ge0$ and $n>s\ge1$.
We call $\cB_{s,t}(\bfx)$ the \emph{$s$-deletion-$t$-transposition ball} centered at $\bfx$.

For a finite set $\cS\subseteq\Sigma_q^{*}$ and integers $s\ge1$ and $t\ge0$, define $\cD_s(\cS)=\cup_{\bfx\in\cS}\cD_s(\bfx)$ and $\cT_{\le t}(\cS)=\cup_{\bfx\in\cS}\cT_{\le t}(\bfx)$. By the definition of $\cB_{s,t}(\bfx)$, it trivially holds that $\cT_{\le t}\parenv{\cD_s(\bfx)},\cD_s\parenv{\cT_{\le t}(\bfx)}\subseteq\cB_{s,t}(\bfx)$. One may ask if $\cB_{s,t}(\bfx)\subseteq \cT_{\le t}\parenv{\cD_s(\bfx)}$ or $\cB_{s,t}(\bfx)\subseteq\cD_s\parenv{\cT_{\le t}(\bfx)}$. The following lemma answers this question.
\begin{lemma}\label{lem_order}
    Let $s\ge 1$ and $t\ge0$ be integers. Let $\bfx\in\Sigma_q^n$, where $n>s$. Then we have $\cB_{s,t}(\bfx)=\cT_{\le t}\parenv{\cD_s(\bfx)}$. Furthermore,
    \begin{enumerate}[$(i)$]
        \item\cite[Lemma 4]{Ryan2018IT} if $q=2$, it holds that $\cT_{\le t}\parenv{\cD_s(\bfx)}=\cD_s\parenv{\cT_{\le t}(\bfx)}$;
        \item if $q>2$, it holds that $\cD_s\parenv{\cT_{\le t}(\bfx)}\subsetneq\cT_{\le t}\parenv{\cD_s(\bfx)}$.
    \end{enumerate}
\end{lemma}
\begin{IEEEproof}
We prove $\cB_{s,t}(\bfx)=\cT_{\le t}\parenv{\cD_s(\bfx)}$ by induction on $s$. Assume that $s=1$.  When $q=2$, it was proved in \cite{Ryan2018IT} that $\cB_{1,t^\prime}(\bfx)=\cT_{\le t^\prime}\parenv{\cD_1(\bfx)}=\cD_1\parenv{\cT_{\le t^\prime}(\bfx)}$ for any $t^\prime$. Therefore, when $s=1$, the conclusion in (i) is true. Following the same argument in the proof of \cite[Lemma 4]{Ryan2018IT}, we can show that $\cD_1\parenv{\cT_{\le t^\prime}(\bfx)}\subseteq\cT_{\le t^\prime}\parenv{\cD_1(\bfx)}$ for any $t^\prime$ when $q>2$. This implies that $\cB_{1,t^\prime}(\bfx)=\cT_{\le t^\prime}\parenv{\cD_1(\bfx)}$.

Now suppose $s\ge 2$ and  $\cB_{s-1,t^\prime}(\bfx)=\cT_{\le t^\prime}\parenv{\cD_{s-1}(\bfx)}$ for any $t^\prime$. Let $\bby\in\cB_{s,t}(\bfx)$. Then there exist sequences $\bbu^{(1)},\ldots,\bbu^{(s)}$, $\bbv^{(1)},\ldots,\bbv^{(s)}$ and non-negative integers $t_1,\ldots,t_{s+1}$ satisfying $t_1+\cdots+t_{s+1}\le t$, such that $\bbu^{(i)}\in\cT_{\le t_i}\parenv{\bbv^{(i-1)}}$ for each $1\le i\le s+1$ and $\bbv^{(i)}\in\cD_1\parenv{\bbu^{(i)}}$ for each $1\le i\le s$. Here, $\bbv^{(0)}=\bfx$ and $\bbu^{(s+1)}=\bby$. Note that $\bbu^{(s)}\in\cB_{s-1,t-t_{s+1}}\parenv{\bfx}$. By induction, we have $\bbu^{(s)}\in\cT_{\le t-t_{s+1}}\parenv{\cD_{s-1}(\bfx)}$. Then it follows that $\bby=\bbu^{(s+1)}\in\cT_{\le t_{s+1}}\parenv{\cD_1\parenv{\bbu^{(s)}}}\subseteq \cT_{\le t_{s+1}}\parenv{\cD_1\parenv{\cT_{\le t-t_{s+1}}\parenv{\cD_{s-1}(\bfx)}}}\subseteq \cT_{\le t_{s+1}}\parenv{\cT_{\le t-t_{s+1}}\parenv{\cD_1\parenv{\cD_{s-1}(\bfx)}}}=\cT_{\le t}\parenv{\cD_s(\bfx)}$. Now we have proved that $\cB_{s,t}(\bfx)\subseteq\cT_{\le t}\parenv{\cD_{s}(\bfx)}$ and thus $\cB_{s,t}(\bfx)=\cT_{\le t}\parenv{\cD_{s}(\bfx)}$. This also implies that $\cD_s\parenv{\cT_{\le t}(\bfx)}\subseteq\cT_{\le t}\parenv{\cD_s(\bfx)}$.

(i) Suppose that $q=2$ and $s\ge2$. It suffices to prove that $\cT_{\le t}\parenv{\cD_s(\bfx)}\subseteq\cD_s\parenv{\cT_{\le t}(\bfx)}$. Let $\bby\in \cT_{\le t}\parenv{\cD_s(\bfx)}$. Then there exist sequences $\bbu^{(1)},\ldots,\bbu^{(s)}$ such that $\bby\in\cT_{\le t}\parenv{\bbu^{(s)}}$ and $\bbu^{(i)}\in\cD_1\parenv{\bbu^{(i-1)}}$ for each $1\le i\le s$. Here, $\bbu^{(0)}=\bfx$. It follows from the discussion in the first paragraph that $\bby\in\cT_{\le t}\parenv{\bbu^{(s)}}\subseteq\cT_{\le t}\parenv{\cD_1\parenv{\bbu^{(s-1)}}}=\cD_1\parenv{\cT_{\le t}\parenv{\bbu^{(s-1)}}}\subseteq\cdots\subseteq\cD_1\parenv{\cD_1\parenv{\cdots\cD_1\parenv{\cT_{\le t}\parenv{\bbu^{(0)}}}}}=\cD_s\parenv{\cT_{\le t}(\bfx)}$. 

(ii) When $q>2$, it is not necessary that $\cT_{\le t}\parenv{\cD_s(\bfx)}\subseteq\cD_s\parenv{\cT_{\le t}(\bfx)}$. For a counterexample, let $\bfx=012$ and $\bby=20\in\cT_{\le1}\parenv{\cD_1(\bfx)}$. Clearly, the deletion of symbol $1$ must occur before the transposition.
\end{IEEEproof}

In this paper, we focus on codes that can correct deletions and multiple transpositions.
\begin{definition}
  Let $\mathcal{C}\subseteq\Sigma_q^n$. If $\cB_{s,t}(\bfx)\cap \cB_{s,t}(\bby)=\emptyset$ for any two distinct sequences $\bfx$ and $\bby$ in $\mathcal{C}$, we call $\mathcal{C}$ an $s$-deletion-$t$-transposition correcting code.
\end{definition}

For a code $\cC\subseteq\Sigma_q^n$ correcting any errors (not necessarily deletions and transpositions), the redundancy of $\cC$ is defined to be $\log\parenv{q^n/\abs{\cC}}$, where $\log(\cdot)$ is the logarithm function with base $2$.

\subsection{General framework for deriving upper bounds}\label{sec_framework}
In \Cref{sec_bounddeltrans,sec_boundextension,sec_boundasy}, we will apply the framework in \cite{Fazeli2015it} to derive upper bounds on the maximum size of specific codes. In this subsection, we briefly describe this framework.

Let $\cH=\parenv{\cV,\cE}$ be a hypergraph, where $\cV=\mathset{v_1,\ldots,v_n}$ is the set of vertices and $\cE=\mathset{E_1,\ldots,E_m}$ is the set of hyperedges. Here, each $E_i$ is a non-empty subset of $\cV$. Define an $n\times m$ matrix $A=(a_{ij})$ as $a_{i,j}=1$ if $v_i\in E_j$ and $a_{i,j}=0$ otherwise.
Let $\cM\subseteq\cE$. If $E\cap E^\prime=\emptyset$ for any distinct $E,E^\prime\in\cM$, we call $\cM$ a matching in $\cH$. Let $\nu\parenv{\cH}$ be the number of hyperedges in the largest matching. It holds that
\begin{equation}\label{eq_framework}
  \nu\parenv{\cH}=\max\mathset{\sum_{j=1}^mz_j:\sum_{j=1}^ma_{ij}z_j\le 1,\forall 1\le i\le n,\text{ where }z_j\in\{0,1\},\forall 1\le j\le m}.  
\end{equation}

Let $\mathscr{C}$ be a channel, which introduces certain errors to sequences passing through it. For $\bfx\in\Sigma_q^n$, let $\cB_{\mathscr{C}}(\bfx)$ be the set of all 
possible outputs when $\bfx$ passes through $\mathscr{C}$. For example, if $\mathscr{C}$ introduces $s$ deletions and at most $t$ transpositions, then $\cB_{\mathscr{C}}(\bfx)=\cB_{s,t}(\bfx)$. Suppose that $\cB_{\mathscr{C}}(\bfx)\subseteq\Sigma_q^m$. We can define a hypergraph $\cH_{\mathscr{C}}=\parenv{\cV,\cE}$, where $\cV=\Sigma_q^{m}$ and $\cE=\mathset{\cB_{\mathscr{C}}(\bfx):\bfx\in\Sigma_q^n}$. Let $\cC\subseteq\Sigma_q^n$ be a code which can correct errors introduced by channel $\mathscr{C}$. Then it is necessary that $\cB_{\mathscr{C}}(\bfx)\cap\cB_{\mathscr{C}}\parenv{\bfx^\prime}$ for any distinct $\bfx,\bfx^\prime\in\cC$. In other words, the set $\mathset{\cB_{\mathscr{C}}(\bfx):\bfx\in\cC}$ is a matching in the hypergraph $\cH_{\mathscr{C}}$.

Next, we define a matrix $A=\parenv{a_{\bby,\bfx}}$, whose rows are indexed by $\Sigma_q^{m}$ and columns are indexed by $\Sigma_q^n$. For $\bfx\in\Sigma_q^n$ and $\bby\in\Sigma_q^{m}$, the entry $a_{\bby,\bfx}$ is given by
$$
a_{\bby,\bfx}=
\begin{cases}
    1,\mbox{ if }\bby\in\cB_{\mathscr{C}}(\bfx),\\
    0,\mbox{ otherwise}.
\end{cases}
$$
By (\ref{eq_framework}), we have
\begin{equation}\label{eq_spherepacking1}
\abs{\cC}\le\max\mathset{\sum_{\bfx\in\Sigma_q^{n}}u_{\bfx}:\sum_{\bfx\in\Sigma_q^n}a_{\bby,\bfx}\cdot u_{\bfx}\le 1,\forall\bby\in\Sigma_q^{m},\text{ where }u_{\bfx}\in\{0,1\},\forall\bfx\in\Sigma_q^{n}}.
\end{equation}
Then it follows from (\ref{eq_spherepacking1}) and \cite[Section II]{Fazeli2015it} that
\begin{align}
        \abs{\cC}&\le\min\mathset{\sum_{\bby\in\Sigma_q^{m}}w_{\bby}:\sum_{\bby\in\Sigma_q^{m}}a_{\bby,\bfx}\cdot w_{\bby}\ge 1,\forall\bfx\in\Sigma_q^{n},\text{ where }w_{\bby}\ge0,\forall\bby\in\Sigma_q^{m}}\notag\\
        &=\min\mathset{\sum_{\bby\in\Sigma_q^{m}}w_{\bby}:\sum_{\bby\in\cB_{\mathscr{C}}(\bfx)}w_{\bby}\ge 1,\forall\bfx\in\Sigma_q^{n},\text{ where }w_{\bby}\ge0,\forall\bby\in\Sigma_q^{m}}\label{eq_spherepacking2}.
\end{align}

It is easy to see that, in this framework, the critical step is to find a suitable $w_{\bby}$ for each $\bby$.

\section{On the Size of $s$-Deletion-$t$-Transposition Balls}\label{sec_ballsize}
This section analyzes the combinatorial properties of  $s$-deletion-$t$-transposition error balls. We begin by characterizing the intersection size between two distinct $1$-transposition balls. Building on this foundation, we establish an exact formula for $1$-deletion-$1$-transposition ball sizes. Subsequently, we derive lower and upper bounds on the size of $1$-deletion-$t$-transposition balls. At last, we bound the size of $s$-deletion-$t$-transposition balls from below and above, for general $s$ and $t$. Beyond revealing some combinatorial properties of $s$-deletion-$t$-transposition balls, results in this section also provide critical guidance for selecting suitable variables $w_{\bby}$, which are needed in the framework described in \Cref{sec_framework}.

\subsection{Intersection of two single-transposition balls}
In this subsection, it is always assumed that $n\ge 2$ is an integer. The next two claims are easy to verify.
\begin{claim}\label{clm_inclusion}
  Let $\bfx,\bby\in\Sigma_q^n$ and $\bfx\ne\bby$. Then $\bfx\in \cT_1(\bby)$ if and only if $\bby\in \cT_1(\bfx)$.
\end{claim}

\begin{claim}\label{clm_matching}
  Let $\bfx,\bby\in\Sigma_q^n$ and $\bfx\ne\bby$. Suppose $T(\bfx,i)=T(\bby,j)$ where $1\le i,j\le n-1$. Then $i\ne j$. Suppose $i<j$. Then $x_{i+1}=y_i$, $x_{j+1}=y_j$ and $x_k=y_k$ when $k\notin\{i,i+1,j,j+1\}$.
\end{claim}

\begin{claim}\label{clm_in}
  Let $\bfx,\bby\in\Sigma_q^n$ and $\bfx\ne\bby$. Suppose $\bfx\in \cT_1(\bby)$. Then $\cT_{\le 1}(\bfx)\cap \cT_{\le 1}(\bby)=\left\{\bfx,\bby\right\}$.
\end{claim}
\begin{IEEEproof}
From \Cref{clm_inclusion} we know that $\left\{\bfx,\bby\right\}\subseteq \cT_{\le 1}(\bfx)\cap \cT_{\le 1}(\bby)$. Next we need to prove that the opposite inclusion is true. Since $\cT_{\le 1}(\bfx)=\{\bfx\}\cup \cT_1(\bfx)$ and $\cT_{\le 1}(\bby)=\{\bby\}\cup \cT_1(\bby)$, it is sufficient to show that $\cT_1(\bfx)\cap \cT_1(\bby)=\emptyset$. Assume on the contrary that $T(\bfx,i)=T(\bby,j)$. Without loss of generality, we can assume $i<j$.

Since $\bfx\in \cT_1(\bby)$, there is some $k$ such that $\bfx=T(\bby,k)$. Then $y_k\ne y_{k+1}$, $x_k=y_{k+1}$, $x_{k+1}=y_k$ and $x_l=y_l$ for all $l\ne k,k+1$. In particular, we have $x_k\ne y_k$ and $x_{k+1}\ne y_{k+1}$. So by \Cref{clm_matching}, we can conclude that $k,k+1\in\{i,j,i+1,j+1\}$ and thus $k\in\{i,i+1,j\}$. If $k=i$, then $j+1\ne k,k+1$ and so $x_{j+1}=y_{j+1}$. On the other hand, \Cref{clm_matching} tells us that $x_{j+1}=y_j$. So we have $y_{j}=y_{j+1}$, which contradicts the fact $y_j\ne y_{j+1}$. If $k=i+1$ or $k=j$, then $i\ne k,k+1$ and so $x_i=y_i$. On the other hand, from \Cref{clm_matching} we know that $x_{i+1}=y_i$. So we have $x_{i}=x_{i+1}$, which is a contradiction. Therefore, we have $\cT_1(\bfx)\cap \cT_1(\bby)=\emptyset$. Now the proof is completed.
\end{IEEEproof}

\begin{claim}\label{clm_notin}
  Let $\bfx,\bby\in\Sigma_q^n$ and $\bfx\ne\bby$. Suppose $\bfx\notin \cT_1(\bby)$. Then $\cT_{\le 1}(\bfx)\cap \cT_{\le 1}(\bby)=\cT_{1}(\bfx)\cap \cT_{1}(\bby)$.
\end{claim}
\begin{IEEEproof}
  The conclusion is clear from \Cref{clm_inclusion} and the fact that $\cT_{\le 1}(\bfx)=\{\bfx\}\cup \cT_1(\bfx)$ and $\cT_{\le 1}(\bby)=\{\bby\}\cup \cT_1(\bby)$.
\end{IEEEproof}

\begin{claim}\label{clm_intersection2}
  Let $\bfx,\bby\in\Sigma_q^n$ and $\bfx\ne\bby$. Suppose $T(\bfx,i_1)=T(\bby,j_1)$, $T(\bfx,i_2)=T(\bby,j_2)$ and $T(\bfx,i_1)\ne T(\bfx,i_2)$. Then $j_2=i_1<j_1=i_2$ or $j_1=i_2<j_2=i_1$. In particular, $\cT_{1}(\bfx)\cap \cT_{1}(\bby)=\left\{T(\bfx,i_1),T(\bfx,i_2)\right\}$ and $|i_1-j_1|=|i_2-j_2|\ge 2$.
\end{claim}
\begin{IEEEproof}
  Since $T(\bfx,i_1)\ne T(\bfx,i_2)$, we have $i_1\ne i_2$ and $j_1\ne j_2$.

  We first assume $i_1<j_1$. Suppose $i_2<j_2$. If $i_1<i_2$, then by \Cref{clm_matching} we have $x_{i_1}=y_{i_1}=x_{i_1+1}$, which is a contradiction. If $i_1>i_2$, by \Cref{clm_matching} we have $x_{i_2}=y_{i_2}=x_{i_2+1}$, which is a contradiction. Therefore, we have $j_2<i_2$. With the help of \Cref{clm_matching}, we further have the following assertions.
  \begin{itemize}
    \item $j_2\ge i_1$. Otherwise, we have $y_{j_2}=x_{j_2}=y_{j_2+1}$, a contradiction.
    \item $j_2\le i_1$. Otherwise, we have $x_{i_1}=y_{i_1}=x_{i_1+1}$, a contradiction.
    \item $j_1\ge i_2$. Otherwise, we have $x_{i_2+1}=y_{i_2+1}=x_{i_2}$, a contradiction.
    \item $j_1\le i_2$. Otherwise, we have $y_{j_1+1}=x_{j_1+1}=y_{j_1}$, a contradiction.
  \end{itemize}
  Therefore, it holds that $j_2=i_1<j_1=i_2$. Similarly, when $i_1>j_1$, we can show that $j_1=i_2<j_2=i_1$. This further implies that $\left|\cT_{1}(\bfx)\cap \cT_{1}(\bby)\right|=2$ and hence $\cT_{1}(\bfx)\cap \cT_{1}(\bby)=\left\{T(\bfx,i_1),T(\bfx,i_2)\right\}$. If $|i_1-j_1|=1$, we have $y_{j_2}=y_{j_2+1}$, which is a contradiction. So $|i_1-j_1|=|i_2-j_2|\ge 2$.
\end{IEEEproof}

By \Cref{clm_matching}, \Cref{clm_in}, \Cref{clm_notin} and \Cref{clm_intersection2}, we have the following theroem.
\begin{theorem}\label{thm_intersection}
Let $\bfx,\bby\in\Sigma_q^n$ and $\bfx\ne\bby$. Then $\left|\cT_{\le 1}(\bfx)\cap \cT_{\le 1}(\bby)\right|\le 2$. Furthermore,
\begin{enumerate}[$(i)$]
    \item $\left|\cT_{\le 1}(\bfx)\cap \cT_{\le 1}(\bby)\right|=2$ if and only if there are some $\bbu,\bbv,\bbw\in\Sigma_q^{*}$, $a,b\in\Sigma_q$, $a^\prime\in\Sigma_q\setminus\{a\}$ and $b^\prime\in\Sigma_q\setminus\{b\}$ such that
        \begin{equation}\label{eq_size21}
          \left\{
          \begin{array}{l}
              \bfx=\bbu aa^\prime\bbv, \\
              \bby=\bbu a^\prime a\bbv,
          \end{array}
          \right.
        \end{equation}
        or
        \begin{equation}\label{eq_size22}
          \left\{
          \begin{array}{l}
              \bfx=\bbu aa^\prime\bbv bb^\prime\bbw, \\
              \bby=\bbu a^\prime a\bbv b^\prime b\bbw.
          \end{array}
          \right.
        \end{equation}
    \item $\left|\cT_{\le 1}(\bfx)\cap \cT_{\le 1}(\bby)\right|=1$ if and only if
    \begin{equation}\label{eq_size23}
      \left\{\bfx,\bby\right\}=\left\{\bbu aa^\prime a^{\prime\prime}\bbv,\bbu a^\prime a^{\prime\prime} a\bbv\right\}
    \end{equation}
    for some $\bbu,\bbv\in\Sigma_q^{*}$, $a\in\Sigma_q$ and $a^\prime, a^{\prime\prime}\in\Sigma_q\setminus\{a\}$.
\end{enumerate}
In particular, if $\cT_{\le 1}(\bfx)\cap \cT_{\le 1}(\bby)\ne\emptyset$ and $\bfx\ne\bby$, then we have
\begin{itemize}
    \item $d_H(\bfx,\bby)\in\{2,4\}$ when $q=2$;
    \item $d_H(\bfx,\bby)\in\{2,3,4\}$ when $q>2$.
\end{itemize}
\end{theorem}
\begin{IEEEproof}
If $\bfx\in \cT_{1}(\bby)$, then $\abs{\cT_{\le 1}(\bfx)\cap \cT_{\le 1}(\bby)}=2$ by \Cref{clm_in}. Now suppose $\bfx\notin \cT_{1}(\bby)$. By \Cref{clm_notin}, we have $\cT_{\le 1}(\bfx)\cap \cT_{\le 1}(\bby)=\cT_{1}(\bfx)\cap \cT_{1}(\bby)$. Then it follows from \Cref{clm_intersection2} that $\abs{\cT_{\le 1}(\bfx)\cap \cT_{\le 1}(\bby)}=2$ if $\abs{\cT_{\le 1}(\bfx)\cap \cT_{\le 1}(\bby)}\ge2$. Now we have proved that $\abs{\cT_{\le 1}(\bfx)\cap \cT_{\le 1}(\bby)}\le 2$ for any distinct $\bfx$ and $\bby$.

(i) Suppose $\abs{\cT_{\le 1}(\bfx)\cap \cT_{\le 1}(\bby)}=2$. If $\bfx\in \cT_{1}(\bby)$, it is easy to see that (\ref{eq_size21}) holds. If $\bfx\notin \cT_{1}(\bby)$, \Cref{clm_notin} implies that there are $i_1,i_2,j_1,j_2$ such that $T(\bfx,i_1)=T(\bby,j_1)$, $T(\bfx,i_2)=T(\bby,j_2)$ and $T(\bfx,i_1)\ne T(\bfx,i_2)$. Now (\ref{eq_size22}) follows from \Cref{clm_intersection2}. On the other hand, if (\ref{eq_size21}) or (\ref{eq_size22}) holds, it is easy to verify that $\abs{\cT_{\le 1}(\bfx)\cap \cT_{\le 1}(\bby)}=2$.

(ii) Suppose $\abs{\cT_{\le 1}(\bfx)\cap \cT_{\le 1}(\bby)}=1$. Then there exist $i,j$, where $i\ne j$, such that $T(\bfx,i)=T(\bby,j)$. If $\abs{i-j}>1$, it holds that $T(\bfx,i),T(\bfx,j)\in \cT_{\le 1}(\bfx)\cap \cT_{\le 1}(\bby)$ and $T(\bfx,i)\ne T(\bfx,j)$. Therefore, it must be that $\abs{i-j}=1$ and we get (\ref{eq_size23}). On the other hand, if (\ref{eq_size23}) holds, it is easy to verify that $\abs{\cT_{\le 1}(\bfx)\cap \cT_{\le 1}(\bby)}=1$.

It is easy to see that $d_H\parenv{\bfx,\bby}=2$ if $\bfx$ and $\bby$ satisfy (\ref{eq_size21}), and $d_H\parenv{\bfx,\bby}=4$ if $\bfx$ and $\bby$ satisfy (\ref{eq_size22}). Now suppose that $\bfx$ and $\bby$ satisfy (\ref{eq_size23}). If $a^\prime=a^{\prime\prime}$, we have $d_H\parenv{\bfx,\bby}=2$. If $a^\prime\ne a^{\prime\prime}$, we have $d_H\parenv{\bfx,\bby}=3$. Particularly, when $q=2$, it must be that $a^\prime=a^{\prime\prime}=1-a$.
\end{IEEEproof}

\subsection{The size of $1$-deletion-$1$-transposition balls}
Let $\bfx\in\Sigma_q^n$ and $1\le i\le j\le n$. The substring $\bfx_{[i,j]}$ is called a \emph{run} if $x_i=x_{i+1}=\cdots=x_j$ and $x_{i-1},x_{j+1}\ne x_i$. Let $r(\bfx)$ denote the number of runs in $\bfx$.
In this and next subsections, denote $r=r(\bfx)$ and write $\bfx$ as $\bfx=a_1^{l_1}\cdots a_r^{l_r}$, where $l_i\ge 1$ and $a_i\ne a_{i+1}$.

Recall that $\cB_{1,1}(\bfx)$ denotes the set of sequences obtained from $\bfx$ by one deletion and at most one transposition from $\bfx$. \Cref{lem_order} says that $\cB_{1,1}(\bfx)=\cup_{\bby\in\cD_1(\bfx)}\cT_{\le 1}(\bby)$. It is well known that $\abs{\cD_1(\bfx)}=r$ and
$\cD_1(\bfx)=\mathset{\bfx^{(i)}:i=1,\ldots,r}$,
where $\bfx^{(i)}\triangleq a_1^{l_1}\cdots a_{i-1}a_{l_i-1}a_{i+1}^{l_{i+1}}\cdots a_r^{l_r}$ is the subsequence obtained by deleting a symbol from the $i$-th run. It follows from the fact $\cB_{1,1}(\bfx)=\cT_{\le 1}\parenv{\cD_1(\bfx)}$ (see \Cref{lem_order}) and the inclusion-exclusion principle that
\begin{equation}\label{eq_exformula}
    \abs{\cB_{1,1}(\bfx)}=\sum_{s=1}^r(-1)^{s-1}\sum_{1\le i_1<\cdots<i_s\le r}\abs{\bigcap_{j=1}^s\cT_{\le1}\parenv{\bfx^{(i_j)}}}.
\end{equation}
Therefore, the problem boils down to calculating $\abs{\bigcap_{j=1}^s\cT_{\le1}\parenv{\bfx^{(i_j)}}}$ for all $s\ge1$. The following trivial lemma will be helpful in our analysis. A proof for the case $q=2$ was given in \cite[Lemma 5]{Abu-Sini2021IT}. This proof also holds when $q>2$.
\begin{lemma}\label{lem_Hammingdistance}
    It holds that $d_H(\bfx^{(i)},\bfx^{(j)})=j-i$ for any $1\le i<j\le r$.
\end{lemma}

\begin{lemma}\label{lem_fourseq}
    For any $1\le i_1<i_2<i_3<i_4\le r$, we have $\bigcap_{j=1}^4\cT_{\le1}\parenv{\bfx^{(i_j)}}=\emptyset$.
\end{lemma}
\begin{IEEEproof}
Suppose on the contrary that $\bigcap_{j=1}^4\cT_{\le1}\parenv{\bfx^{(i_j)}}\ne\emptyset$.
    Recall from \Cref{thm_intersection} that $d_H(\bfx^{(i)},\bfx^{(j)})\in\{2,3,4\}$ if $\cT_{\le 1}(\bfx^{(i)})\cap \cT_{\le 1}(\bfx^{(i)})\ne\emptyset$ and $i\ne j$. Then it follows from \Cref{lem_Hammingdistance} that $i_2-i_1,i_3-i_2,i_4-i_3,i_4-i_1\in\{2,3,4\}$, which is impossible.
\end{IEEEproof}

This lemma says that $\abs{\bigcap_{j=1}^s\cT_{\le1}\parenv{\bfx^{(i_j)}}}=0$ for all $s\ge 4$. It remains to study cases $s\in\{1,2,3\}$. The case $s=1$ is easy to handle with the help of the next lemma, whose proof is clear from the definition of single-transposition balls.
\begin{lemma}\label{lem_transpositionballsize}
    For any sequence $\bbu$, it holds that $\abs{\cT_{\le1}(\bbu)}=r(\bbu)$.
\end{lemma}

This lemma motivates the following definition of multisets of different runs in $\bfx$.
\begin{definition}\label{dfn_setofruns}
    Let $\bfx\in\Sigma_q^n$. Recall that we write $\bfx$ as $\bfx=a_1^{l_1}\cdots a_r^{l_r}$, where $r=r(\bfx)$ and $a_i^{l_i}$ ($i=1,\ldots,r$) are all runs in $\bfx$. Define the following \emph{multisets} of runs in $\bfx$:
\begin{equation}
\begin{array}{l}
    \cR_{1}^{\prime}=\mathset{a_i^{l_i}:1<i<r,l_i=1,a_{i-1}=a_{i+1}},\\
    \cR_1^{side}=\mathset{a_i^{l_i}:l_i=1,i= 1\text{ or }r},\\
    \cR_{\ge2}=\mathset{a_i^{l_i}:1\le i\le r,l_i\ge2}.
\end{array}
\end{equation}
When $q>2$, we further define
\begin{equation*}
    \cR_1^{\prime\prime}=\mathset{a_i^{l_i}:1<i<r,l_i=1,a_{i-1}\ne a_{i+1}}.
\end{equation*}
\end{definition}
Clearly, these multisets depend on specific $\bfx$. We omit $\bfx$ in the definition since $\bfx$ will be clear from the context.
With notations in \Cref{dfn_setofruns}, we show that $\abs{\cT_{\le1}\parenv{\bfx^{(i)}}}$ is determined by $r$ and which multiset $a_i^{l_i}$ belongs to.
\begin{corollary}\label{cor_oneseq}
    Let $\bfx$ and $\bfx^{(i)}$ be as above. Then it holds that
    \begin{equation*}
        \abs{\cT_{\le1}\parenv{\bfx^{(i)}}}=
        \begin{cases}
            r,\mbox{ if }a_i^{l_i}\in\cR_{\ge2},\\
            r-2,\mbox{ if }a_i^{l_i}\in\cR_{1}^{\prime},\\
            r-1,\mbox{ if }a_i^{l_i}\in\cR_{1}^{\prime\prime},\\
            r-1,\mbox{ if }a_i^{l_i}\in\cR_{1}^{side}.
        \end{cases}
    \end{equation*}
\end{corollary}
\begin{IEEEproof}
    The conclusion follows from \Cref{lem_transpositionballsize} and the fact that
    \begin{equation*}
        r\parenv{\bfx^{(i)}}=
        \begin{cases}
            r,\mbox{ if }a_i^{l_i}\in\cR_{\ge2},\\
            r-2,\mbox{ if }a_i^{l_i}\in\cR_{1}^{\prime},\\
            r-1,\mbox{ if }a_i^{l_i}\in\cR_{1}^{\prime\prime},\\
            r-1,\mbox{ if }a_i^{l_i}\in\cR_{1}^{side}.
        \end{cases}
    \end{equation*}
\end{IEEEproof}

Next, we analyze the case $s=2$.
\begin{lemma}\label{lem_twoseq}
    For $1\le i<j\le r$, it holds that
    \begin{enumerate}[$(i)$]
        \item $\bfx^{(i)}$ and $\bfx^{(j)}$ satisfy (\ref{eq_size21}) if and only if $j=i+2$, $l_{i+1}=1$ and $a_i=a_{i+2}$ (when $q=2$, it must be that $a_i=a_{i+2}$);
        \item $\bfx^{(i)}$ and $\bfx^{(j)}$ satisfy (\ref{eq_size22}) if and only if $j=i+4$ $l_{i+1}=l_{i+3}=1$ and $a_i=a_{i+2}=a_{i+4}$ (when $q=2$, it must be that $a_i=a_{i+2}=a_{i+4}$);
        \item $\bfx^{(i)}$ and $\bfx^{(j)}$ satisfy (\ref{eq_size23}) if and only if $j=i+2$, $l_{i+1}=2$ and $a_i=a_{i+2}$ (when $q=2$, it must be that $a_i=a_{i+2}$).
    \end{enumerate}
\end{lemma}
\begin{IEEEproof}
The first two claims follow trivially from definitions of $\bfx^{(i)}$ and $\bfx^{(j)}$ and \Cref{lem_Hammingdistance}. Similarly, the third claim is true if we can show that $d_H\parenv{\bfx^{(i)},\bby^{(j)}}\ne3$ when $q>2$ and $\bfx^{(i)}$ and $\bfx^{(j)}$ satisfy (\ref{eq_size23}). Suppose on the contrary that $d_H\parenv{\bfx^{(i)},\bby^{(j)}}=3$. Then $j-i=3$ and
\begin{equation*}
    \begin{array}{l}
         \bfx^{(i)}=\cdots a_{i}^{l_i-1}a_{i+1}a_{i+1}^{l_{i+1}-1}a_{i+2}a_{i+2}^{l_{i+2}-1}a_{i+3}a_{i+3}^{l_{i+3}-1}\cdots,  \\
         \bfx^{(j)}=\cdots a_{i}^{l_i-1}a_{i}\quad a_{i+1}^{l_{i+1}-1}a_{i+1}a_{i+2}^{l_{i+2}-1}a_{i+2}a_{i+3}^{l_{i+3}-1}\cdots .
    \end{array}
\end{equation*}
Since $\bfx^{(i)}$ and $\bfx^{(j)}$ satisfy (\ref{eq_size23}), it is necessary that $l_{i+1}=l_{i+2}=1$ and $a_{i+1}=a_{i+2}$, which is a contradiction.
\end{IEEEproof}

Now we consider the case $s=3$.
\begin{lemma}\label{lem_threeseq}
    Suppose that $1\le i<j<k\le r$. It holds that $\cT_{\le1}\parenv{\bfx^{(i)}}\cap\cT_{\le1}\parenv{\bfx^{(j)}}\cap\cT_{\le1}\parenv{\bfx^{(k)}}\ne\emptyset$ if and only if $j-i=k-j=2$, $l_{i+1}=l_{i+3}=1$ and $a_{i}=a_{i+2}=a_{i+4}$ (when $q=2$, it must be that $a_i=a_{i+2}=a_{i+4}$). In this case, we have $\abs{\cT_{\le1}\parenv{\bfx^{(i)}}\cap\cT_{\le1}\parenv{\bfx^{(j)}}\cap\cT_{\le1}\parenv{\bfx^{(k)}}}=1$.
\end{lemma}
\begin{IEEEproof}
    We first prove the ``$\Leftarrow$" direction. Suppose that $j-i=k-j=2$ and $l_{i+1}=l_{i+3}=1$. Then we have
    \begin{equation*}
    \begin{array}{l}
         \bfx^{(i)}=\cdots a_i^{l_i-1}a_{i+1}a_{i+2}a_{i+2}^{l_{i+2}-1}a_{i+3}a_{i+4}a_{i+4}^{l_{i+4}-1}\cdots,\\
        \bfx^{(j)}=\cdots a_i^{l_i-1}a_ia_{i+1}a_{i+2}^{l_{i+2}-1}a_{i+3}a_{i+4}a_{i+4}^{l_{i+4}-1}\cdots,\\
        \bfx^{(k)}=\cdots a_i^{l_i-1}a_ia_{i+1}a_{i+2}^{l_{i+2}-1}a_{i+2}a_{i+3}a_{i+4}^{l_{i+4}-1}\cdots.
    \end{array}
    \end{equation*}
Since $a_i=a_{i+2}=a_{i+4}$, it is easy to verify that
$$
\cT_{\le1}\parenv{\bfx^{(i)}}\cap\cT_{\le1}\parenv{\bfx^{(j)}}\cap\cT_{\le1}\parenv{\bfx^{(k)}}=\mathset{\cdots a_i^{l_i-1}a_ia_{i+1}a_{i+2}^{l_{i+2}-1}a_{i+3}a_{i+4}a_{i+4}^{l_{i+4}-1}\cdots}.
$$

Next, we prove the ``$\Rightarrow$" direction. Suppose that $\cT_{\le1}\parenv{\bfx^{(i)}}\cap\cT_{\le1}\parenv{\bfx^{(j)}}\cap\cT_{\le1}\parenv{\bfx^{(k)}}\ne\emptyset$. Then \Cref{thm_intersection} and \Cref{lem_Hammingdistance} imply that $j-i,k-j,k-i\in\{2,3,4\}$. Therefore, we have $j-i=k-j=2$. Then it follows that
\begin{equation*}
    \begin{array}{l}
         \bfx^{(i)}=\cdots a_i^{l_i-1}a_{i+1}^{l_{i+1}}a_{i+2}a_{i+2}^{l_{i+2}-1}a_{i+3}^{l_{i+3}}a_{i+4}a_{i+4}^{l_{i+4}-1}\cdots,\\
        \bfx^{(k)}=\cdots a_i^{l_i-1}a_ia_{i+1}^{l_{i+1}}a_{i+2}^{l_{i+2}-1}a_{i+2}a_{i+3}^{l_{i+3}}a_{i+4}^{l_{i+4}-1}\cdots.
    \end{array}
    \end{equation*}
    The assumption that $\cT_{\le1}\parenv{\bfx^{(i)}}\cap\cT_{\le1}\parenv{\bfx^{(k)}}\ne\emptyset$ implies that $\bfx^{(i)}$ and $\bfx^{(k)}$ must satisfy one of (\ref{eq_size21}), (\ref{eq_size22}) and (\ref{eq_size23}). Since $d_H\parenv{\bfx^{(i)},\bfx^{(k)}}=4$, the two sequences
    $\bfx^{(i)}$ and $\bfx^{(k)}$ can not satisfy (\ref{eq_size21}) or (\ref{eq_size23}). On the other hand, $\bfx^{(i)}$ and $\bfx^{(k)}$ satisfy (\ref{eq_size22}) if and only if $l_{i+1}=l_{i+3}=1$ and $a_{i}=a_{i+2}=a_{i+4}$.
\end{IEEEproof}

The following definition draws inspiration from \Cref{cor_oneseq} and \Cref{lem_threeseq,lem_twoseq}.
\begin{definition}
   Let $n\ge2$ be an integer. For a sequence $\bfx\in\Sigma_q^n$, let $r=r(\bfx)$ and write $\bfx$ as $\bfx=a_1^{l_1}\cdots a_r^{l_r}$, where $l_i\ge 1$ and $a_i\ne a_{i+1}$. Let $\cR_{\ge2}$, $\cR_{1}^{\prime}$, $\cR_{1}^{\prime\prime}$ and $\cR_{1}^{side}$ be defined as in \Cref{dfn_setofruns}. Define
\begin{equation*}
\begin{array}{l}
r_1^{\prime}=\abs{\cR_{1}^{\prime}}=\abs{\mathset{1<i<r:l_i=1,a_{i-1}=a_{i+1}}},\\
    r_1^{\prime\prime}=\abs{\cR_{1}^{\prime\prime}}=\abs{\mathset{1<i<r:l_i=1,a_{i-1}\ne a_{i+1}}},\\
    r_1^{side}=\abs{\cR_1^{side}}=\abs{\mathset{i:l_i=1,i= 1\text{ or }r}},\\
    r_{\ge2}=\abs{\cR_{\ge2}}=\abs{\mathset{1\le i\le r:l_i\ge2}},\\
    r_1^{pair}=\abs{\mathset{1\le i\le r-4:l_{i+1}=l_{i+3}=1,a_{i}=a_{i+2}=a_{i+4}}},\\
    r_2^{in}=\abs{\mathset{1\le i\le r-2:l_{i+1}=2,a_{i}=a_{i+2}}}.
\end{array}
\end{equation*}
\end{definition}

The following theorem establishes that the size of $\cB_{1,1}(\bfx)$ is completely determined by the aforementioned parameters.
\begin{theorem}\label{thm_DTballsize}
    For a sequence $\bfx\in\Sigma_q^n$, let $r$, $r_{1}^{\prime}$, $r_{1}^{\prime\prime}$, $r_{1}^{side}$, $r_{\ge2}$, $r_1^{pair}$ and $r_2^{in}$ be as above. Then we have
    \begin{equation}\label{eq_DTballsize}
        \abs{\cB_{1,1}(\bfx)}=r^2-4r_{1}^{\prime}-r_1^{\prime\prime}-r_1^{side}-r_1^{pair}-r_2^{in}.
    \end{equation}
    This implies that
    \begin{equation*}
        \abs{\cB_{1,1}(\bfx)}\ge
        \begin{cases}
            \max\{r(r-1),1\},\mbox{ if }r_1^\prime=0,\\
            r(r-2),\mbox{ if }r_1^\prime=1,\\
            r(r-5)+9,\mbox{ if }r_1^\prime\ge2.
        \end{cases}
    \end{equation*}
    In particular, we have $\abs{\cB_{1,1}(\bfx)}\ge r(r-5)+9$ whenever $r\ge3$.
\end{theorem}
\begin{IEEEproof}
By \Cref{cor_oneseq}, we have
\begin{equation}\label{eq_part1}
  \sum_{i=1}^{r}\abs{\cT_{\le1}\parenv{\bfx^{(i)}}}=r\cdot r_{\ge2}+(r-2)r_{1}^{\prime}+(r-1)\parenv{r_1^{side}+r_1^{\prime\prime}}.
\end{equation}
By \Cref{thm_intersection} and \Cref{lem_twoseq}, we have
\begin{equation}\label{eq_part2}
    \sum_{1\le i_1<i_2\le r}\abs{\cT_{\le1}\parenv{\bfx^{(i_1)}}\cap\cT_{\le1}\parenv{\bfx^{(i_2)}}}=2r_{1}^{\prime}+2r_1^{pair}+r_2^{in}.
\end{equation}
By \Cref{thm_intersection} and \Cref{lem_threeseq}, we have
\begin{equation}\label{eq_part3}
    \sum_{1\le i_1<i_2<i_3\le r}\abs{\bigcap_{j=1}^3\cT_{\le1}\parenv{\bfx^{(i_j)}}}=r_1^{pair}.
\end{equation}
Now \Cref{eq_DTballsize} follows from \Cref{eq_exformula,eq_part1,eq_part2,eq_part3}, \Cref{lem_fourseq} and the fact that $r=r_{\ge2}+r_{1}^{\prime}+r_{1}^{\prime\prime}+r_1^{side}$.

By definition, we have $r_1^{pair}\le \max\mathset{r_{1}^{\prime}-1,0}$. When $r_1^\prime=0$, it follows from (\ref{eq_DTballsize}) that $\abs{\cB_{1,1}(\bfx)}-r(r-1)=r_{\ge2}-r_2^{in}\ge0$. Also, note that $\abs{\cB_{1,1}(\bfx)}\ge1$ for any $\bfx$. 
Next, suppose that $r_1^\prime=1$. Then we have $r\ge3$. Therefore, we have $r_{\ge2}\ge2-r_1^{side}$. Combining this with (\ref{eq_DTballsize}), we obtain $\abs{\cB_{1,1}(\bfx)}-r(r-2)=r_1^{\prime\prime}+r_1^{side}+2r_{\ge2}-r_2^{in}-2\ge r_1^{side}+r_{\ge2}+r_{\ge2}-r_2^{in}-2\ge r_{\ge2}-r_2^{in}\ge0$.

At last, suppose $r_1^{\prime}\ge2$. Since $r_1^{pair}\le r_1^{\prime}-1$, it follows from (\ref{eq_DTballsize}) that $\abs{\cB_{1,1}(\bfx)}\ge r^2-5r_{1}^{\prime}-r_1^{\prime\prime}-r_1^{side}-r_2^{in}+1$. Then it follows that $\abs{\cB_{1,1}(\bfx)}-r(r-5)-1\ge 4r_{1}^{side}+4r_{1}^{\prime\prime}+5r_{\ge2}-r_2^{in}\ge 4r_1^{side}+4r_{\ge2}+r_{\ge2}-r_2^{in}\ge 4r_1^{side}+8-4r_1^{side}+r_{\ge2}-r_2^{in}\ge 8$. Here, we also use the fact $r_{\ge2}\ge2-r_1^{side}$. Now the proof is completed.
\end{IEEEproof}

\begin{example}
 Let $\bfx=0201001$. It is easy to see that $r=6$, $r_1^\prime=r_1^{side}=2$ and $r_1^{\prime\prime}=r_1^{pair}=r_2^{in}=1$. By definition we have $\bfx^{(1)}=201001$, $\bfx^{(2)}=001001$, $\bfx^{(3)}=021001$, $\bfx^{(4)}=020001$, $\bfx^{(5)}=020101$ and $\bfx^{(6)}=020100$. Furthermore, we obtain
 \begin{equation*}
     \begin{array}{l}
     \cT_{\le1}\parenv{\bfx^{(1)}}=\mathset{201001,021001,210001,200101,201010},\\
     \cT_{\le1}\parenv{\bfx^{(2)}}=\mathset{001001,010001,000101,001010},\\
     \cT_{\le1}\parenv{\bfx^{(3)}}=\mathset{\bm{021001},\bm{201001},012001,020101,021010},\\
     \cT_{\le1}\parenv{\bfx^{(4)}}=\mathset{020001,200001,002001,020010},\\
     \cT_{\le1}\parenv{\bfx^{(5)}}=\mathset{\bm{020101},\bm{200101},002101,\bm{021001},020011,020110},\\
     \cT_{\le1}\parenv{\bfx^{(6)}}=\mathset{020100,200100,002100,021000,\bm{020010}},
     \end{array}
 \end{equation*}
 where sequences that appear not for the first time are marked in bold font.
 Removing repeated sequences, we obtain $\abs{\cB_{1,1}(\bfx)}=\abs{\cup_{i=1}^4\cT_{\le1}\parenv{\bfx^{(i)}}}=23=r^2-4r_{1}^{\prime}-r_1^{\prime\prime}-r_1^{side}-r_1^{pair}-r_2^{in}$. This verifies \Cref{eq_DTballsize}.
\end{example}

\subsection{Bounds on the size of $1$-deletion-$t$-transposition balls}
We now investigate the size of $\cB_{1,t}(\bfx)$ for general $t$. According to \Cref{lem_order}, we have $\cB_{1,t}(\bfx)=\cup_{i=1}^{r}\cT_{\le t}\parenv{\bfx^{(i)}}$. One may try to follow the same idea in previous subsection to calculate $\abs{\cB_{1,t}(\bfx)}$. However, it is not an easy task to calculate $\abs{\bigcap_{j=1}^s\cT_{\le t}\parenv{\bfx^{(i_j)}}}$ for general $t\ge2$. Therefore, instead of giving an exact formula for $\abs{\cB_{1,t}(\bfx)}$, we aim to bound the size of $1$-deletion-$t$-transposition balls.

Recall that in the definition of $\cT_{t}(\bfx)$, the $t$ adjacent transpositions may occur sequentially. To get a lower bound on $\abs{\cT_{t}(\bfx)}$, it is convenient to consider the case where all $t$ transpositions occur simultaneously. Let $\cT_{t}^{\prime}(\bfx)$ be the set of all sequences obtained from $\bfx$ by \emph{exactly} $t$ \emph{simultaneous} transpositions. In other words, the $t$ transposed pairs do not overlap with each other. It is clear that $\cT_{t}^\prime(\bfx)\subseteq\cT_{t}(\bfx)$.

For each $\bby\in\cT_t^\prime(\bfx)$, there exist $k_1,\ldots,k_t$ satisfying $k_{i+1}-k_i>2$, such that $\bby$ is obtained from $\bfx$ by transposing $x_{k_i}$ and $x_{k_{i+1}}$ for all $1\le i\le t$. For this reason, we also write $\bby=T_{k_1,\ldots,k_t}(\bfx)$. The next lemma will be helpful in the proof of \Cref{lem_lowerboundttransposition,lem_asyballsize}.
\begin{lemma}\label{lem_transdistinct}
    Let $\bfx\in\Sigma_q^n$. Suppose that integers $1\le k_1,\ldots,k_t,l_1,\ldots,l_t<n$ satisfy $k_{i+1}-k_i,l_{i+1}-l_i>2$ for all $i$ and $x_{k_i}\ne x_{k_{i+1}}$, $x_{l_i}\ne x_{l_{i+1}}$ for all $i$. If $\parenv{k_1,\ldots,k_t}\ne\parenv{l_1,\ldots,l_t}$, then $T_{k_1,\ldots,k_t}(\bfx)\ne T_{l_1,\ldots,l_t}(\bfx)$.
\end{lemma}
\begin{IEEEproof}
    Let $\bby=T_{k_1,\ldots,k_t}(\bfx)$ and $\bbz=T_{l_1,\ldots,l_t}(\bfx)$. Both $\bby$ and $\bbz$ are obtained from $\bfx$ by altering exactly $2t$ 
    positions. Note that each adjacent transposition alters exactly two positions. Therefore, if $\parenv{k_1,\ldots,k_t}\ne\parenv{l_1,\ldots,l_t}$, there must be some $i$, such that at least one of $x_{k_i}$ and $x_{k_{i+1}}$ is not affected by the $t$ simultaneous transpositions at positions $l_1,\ldots,l_{t}$. This implies that $y_{k_i}\ne z_{k_i}$ or $y_{k_{i+1}}\ne z_{k_{i+1}}$. Now the proof is completed.
\end{IEEEproof}

By convention, let $\binom{m}{n}$ denote the binomial coefficients, where $m,n$ are integers and $0\le n\le m$. In addition, we set $\binom{m}{0}=1$ for any integer $m$, and $\binom{m}{n}=0$ for other values of $m$ and $n$.
Regarding the size of $\cT_{t}^\prime(\bfx)$, we have the following lemma, which is implicit in the proof of \cite[Theorem 3.1]{Mladen2024researchgate}. Since \cite{Mladen2024researchgate} is not a peer-reviewed work, we present here the proof implied in \cite{Mladen2024researchgate} for readers to verify.
\begin{lemma}\cite{Mladen2024researchgate}\label{lem_lowerboundttransposition}
    Let $\bfx\in\Sigma_q^n$ be a sequence with $r$ runs, where $r\ge 2t+1$. Then it holds that
    \begin{equation}\label{eq_simultran1}
        \abs{\cT_{t}^{\prime}(\bfx)}\ge\sum_{i=0}^{t}\binom{\floorenv{\frac{r}{2}}}{i}\binom{\floorenv{\frac{r}{2}}-2i-1}{t-i}.
    \end{equation}
    In particular, we have $\abs{\cT_{t}^{\prime}(\bfx)}\ge\binom{\floorenv{r/2}}{t}\ge\parenv{\frac{r-1}{2t}}^t$.
\end{lemma}
\begin{IEEEproof}
Recall that each sequence $\bfx\in\Sigma_q^n$ can be written as $\bfx=a_1^{l_1}\cdots a_r^{l_r}$, where $a_1^{l_1},\ldots,a_r^{l_r}$ are all distinct runs in $\bfx$. If $i$ is odd (or even), we call $a_i^{l_i}$ an odd-numbered (or even-numbered) run. Denote $I_{o}=\mathset{i:1\le i\le r,i\text{ is odd}}$. Clearly, we have $\abs{I_o}\ge\floorenv{r/2}$.

Note that each of the $t$ simultaneous transpositions occurs either at the left or at the right boundary of an odd-numbered run. 
We can choose $t$ simultaneous transpositions in the following way. For each $0\le u\le\floorenv{r/2}$, choose $i_1,\ldots,i_u\in I_o$. For each $1\le k\le u$, transpose $a_{i_k}$ and $a_{i_{k}+1}$. In other words, we transpose the right-most symbol in the $i_k$-th run with the left-most symbol in the $(i_k+1)$-th run. 
Next, choose $j_1,\ldots,j_{t-u}\in I_o\setminus\parenv{\mathset{i_1,\ldots,i_u}\cup\mathset{i_1+2,\ldots,i_u+2}\cup\{1\}}$. For each $1\le k\le t-u$, transpose $a_{j_k}$ and $a_{j_k-1}$. In other words,  we transpose the left-most symbol in the $j_k$-th run with the right-most symbol in the $(j_k-1)$-th run.

It is clear from the choice of $\parenv{i_1,\ldots,i_u,j_1,\ldots,j_{t-u}}$ that the $t$ transposed pairs are mutually non-overlapping. According to \Cref{lem_transdistinct}, different choices of $\parenv{i_1,\ldots,i_u,j_1,\ldots,j_{t-u}}$ result in different sequences in $\cT_t^\prime(\bfx)$. Then the proof of (\ref{eq_simultran1}) is completed by noticing that $\abs{\mathset{i_1,\ldots,i_u}\cup\mathset{i_1+2,\ldots,i_u+2}\cup\{1\}}\le 2u+1$. The second lower bound on $\cT_t^\prime(\bfx)$ follows by assigning $i=t$ and the fact that $\binom{m}{k}\ge(m/k)^k$ for any $m\ge k\ge 1$.
\end{IEEEproof}

\begin{theorem}\label{thm_bound1delttrans}
    Let $t\ge1$ be an integer and $\bfx\in\Sigma_q^n$ be a sequence with $r$ runs, where $r\ge 8t+3$. Then we have
    \begin{equation*}
        r\parenv{\frac{r-4t-3}{4t}}^t\le r\sum_{i=0}^{t}\binom{\floorenv{\frac{r-4t-1}{4}}}{i}\binom{\floorenv{\frac{r-4t-1}{4}}-2i-1}{t-i}\le\abs{\cB_{1,t}(\bfx)}\le r^2 \prod_{i=1}^{t-1}(r+2i).
    \end{equation*}
\end{theorem}
\begin{IEEEproof}
    Recall that $\cD_1(\bfx)=\mathset{\bfx^{(i)}:i=1,\ldots,r}$, where $\bfx^{(i)}=a_1^{l_1}\cdots a_{i-1}^{l_{i-1}}a_i^{l_i-1}a_{i+1}^{l_{i+1}}\cdots a_r^{l_r}$ is obtained from $\bfx$ by deleting one symbol in the $i$-th run. Then $\cB_{1,t}(\bfx)=\cup_{i=1}^{r}\cT_{\le t}\parenv{\bfx^{(i)}}$. Note that a transposition increases the number of runs by at most two. Then it follows from \Cref{lem_transpositionballsize} that $\abs{\cT_{\le t}\parenv{\bfx^{(i)}}}\le\prod_{i=0}^{t-1}\parenv{r\parenv{\bfx^{(i)}}+2i}\le\prod_{i=0}^{t-1}\parenv{r+2i}$. Therefore, we have $\abs{\cB_{1,t}(\bfx)}\le\sum_{i=1}^r\abs{\cT_{\le t}\parenv{\bfx^{(i)}}}\le r^2 \prod_{i=1}^{t-1}(r+2i)$. This proves the upper bound.
    
   To derive the lower bound, we seek for a subset $\cS_i\subseteq\cT_{\le t}\parenv{\bfx^{(i)}}$ for each $i$, such that $\cS_i\cap\cS_j=\emptyset$ whenever $i\ne j$. Then it follows that $\abs{\cB_{1,t}(\bfx)}\ge\sum_{i=1}^r\abs{\cS_i}$.
    Suppose $1\le i<j\le r$. By definition, we have
    \begin{equation}\label{eq_comparison}
        \begin{aligned}
            \bfx^{(i)}=\cdots a_{i}^{l_i-1}a_{i+1}^{l_{i+1}}\cdots a_{j-1}^{l_{j-1}}a_j a_{j}^{l_j-1}\cdots,\\
            \bfx^{(j)}=\cdots a_{i}^{l_i-1}a_{i}a_{i+1}^{l_{i+1}}\cdots a_{j-1}^{l_{j-1}} a_{j}^{l_j-1}\cdots.
        \end{aligned}
    \end{equation}
This implies that
\begin{equation}\label{eq_partequal}
    \begin{aligned}
        \bfx^{(i)}_{\sparenv{1,l_1+\cdots+l_{i-1}}}=a_1^{l_1}\cdots a_{i-1}^{l_{i-1}}=\bfx^{(j)}_{\sparenv{1,l_1+\cdots+l_{i-1}}},\\
        \bfx^{(i)}_{\sparenv{l_1+\cdots+l_{j},n}}=a_{j+1}^{l_{j+1}}\cdots a_{r}^{l_{r}}=\bfx^{(j)}_{\sparenv{l_1+\cdots+l_{j},n}},
    \end{aligned}
\end{equation}
and $d_H\parenv{\bfx^{(i)},\bfx^{(j)}}=d_H\parenv{a_{i+1}^{l_{i+1}}\cdots a_{j-1}^{l_{j-1}}a_j,a_{i}a_{i+1}^{l_{i+1}}\cdots a_{j-1}^{l_{j-1}}}=j-i$. Note that $d_H(\bbu,\bbv)\le 2$ if $\bbu\in\cT_{\le 1}(\bbv)$. Therefore, when $j-i\ge 4t+1$, we have $\cT_{\le t}\parenv{\bfx^{(i)}}\cap \cT_{\le t}\parenv{\bfx^{(j)}}=\emptyset$.
    
The above discussion inspires the definition of $\cS_i$'s. For each $1\le i\le r$, define
    $$
    \cS_i=
    \begin{cases}
        a_1^{l_1}\cdots a_{i}^{l_i-1}\cT_{t}^{\prime}\parenv{a_{i+1}^{l_{i+1}}\cdots a_r^{l_r}},\mbox{ if }i\le 4t+1,\\
       \cT_{t}^{\prime}\parenv{a_1^{l_1}\cdots a_{i-4t-1}^{l_{i-4t-1}}}a_{i-4t}^{l_{i-4t}}\cdots a_{i}^{l_i-1}a_{i+1}^{l_{i+1}}\cdots a_r^{l_r}\cup a_1^{l_1}\cdots a_{i}^{l_i-1}\cT_{t}^{\prime}\parenv{a_{i+1}^{l_{i+1}}\cdots a_r^{l_r}} ,\mbox{ if }i\ge 4t+2.
    \end{cases}
    $$
In other words, when $i\le 4t+1$, all the $t$ transpositions are applied to the substring $a_{i+1}^{l_{i+1}}\cdots a_r^{l_r}$ of $\bfx^{(i)}$ simultaneously. When $t\ge 4t+2$, all the $t$ transpositions are applied to the substring $a_1^{l_1}\cdots a_{i-4t-1}^{l_{i-4t-1}}$ simultaneously, or to the substring $a_{i+1}^{l_{i+1}}\cdots a_r^{l_r}$ simultaneously. Clearly, we have $\cS_i\subseteq\cT_{t}\parenv{\bfx^{(i)}}$.
Since there are at least $(r-4t-1)/2$ runs in the substring $a_1^{l_1}\cdots a_{i-4t-1}^{l_{i-4t-1}}$ or there are at least $(r-4t-1)/2$ runs in the substring $a_{i+1}^{l_{i+1}}\cdots a_r^{l_r}$, it follows from \Cref{lem_lowerboundttransposition} that $\abs{\cS_i}\ge\sum_{i=0}^{t}\binom{\floorenv{(r-4t-1)/4}}{i}\binom{\floorenv{(r-4t-1)/4}-2i-1}{t-i}\ge\parenv{\frac{r-4t-3}{4t}}^t$.

It remains to show that $\cS_i\cap\cS_j=\emptyset$ when $i<j$. 
Suppose on the contrary that there are some $i$ and $j$ with $i<j$, such that $\cS_i\cap\cS_j\ne\emptyset$. According to the discussion immediately after (\ref{eq_partequal}), we can further assume that $j-4t\le i$. Let $\bbz\in\cS_i\cap\cS_j$. It follows from the definition of $\cS_j$ that (i) $\bbz\in a_1^{l_1}\cdots a_{j}^{l_j-1}\cT_{t}^{\prime}\parenv{a_{j+1}^{l_{j+1}}\cdots a_r^{l_r}}$, or (ii) $\bbz\in \cT_{t}^{\prime}\parenv{a_1^{l_1}\cdots a_{j-4t-1}^{l_{j-4t-1}}}a_{j-4t}^{l_{j-4t}}\cdots a_{j}^{l_j-1}a_{j+1}^{l_{j+1}}\cdots a_r^{l_r}$.

We firstly consider case (i). In this case, $\bbz$ is obtained from $\bfx^{(j)}$ by applying all $t$ transpositions in the substring $\bfx^{(j)}_{\sparenv{l_1+\cdots+l_{j},n}}$. Since either in $\bfx^{(i)}$ or $\bfx^{(j)}$, the $t$ transposed pairs do not overlap and $\bfx^{(i)}_{\sparenv{l_1+\cdots+l_{j},n}}=\bfx^{(j)}_{\sparenv{l_1+\cdots+l_{j},n}}$, we can conclude that $\bbz$ is obtained from $\bfx^{(i)}$ by applying all $t$ transpositions in the substring $\bfx^{(i)}_{\sparenv{l_1+\cdots+l_{j},n}}$. By (\ref{eq_comparison}), this implies that $a_{i+1}^{l_{i+1}}\cdots a_{j-1}^{l_{j-1}}a_j=\bfx^{(i)}_{\sparenv{l_1+\cdots+l_i,l_1+\cdots+l_{j-1}}}=\bbz_{\sparenv{l_1+\cdots+l_i,l_1+\cdots+l_{j-1}}}=\bfx^{(j)}_{\sparenv{l_1+\cdots+l_i,l_1+\cdots+l_{j-1}}}=a_{i}a_{i+1}^{l_{i+1}}\cdots a_{j-1}^{l_{j-1}}$, which is a contradiction.

Now we consider case (ii). In this case, $\bbz$ is obtained from $\bfx^{(j)}$ by applying all $t$ transpositions in the substring $\bfx^{(j)}_{\sparenv{1,l_1+\cdots+l_{j-4t-1}}}$. By the assumption that $j-4t\le i$, we have $j-4t-1\le i-1$. Then it follows from (\ref{eq_comparison}) that $\bfx^{(i)}_{\sparenv{1,l_1+\cdots+l_{j-4t-1}}}=\bfx^{(j)}_{\sparenv{1,l_1+\cdots+l_{j-4t-1}}}$. Therefore, $\bbz$ is obtained from $\bfx^{(i)}$ by applying all $t$ transpositions in the substring $\bfx^{(i)}_{\sparenv{1,l_1+\cdots+l_{j-4t-1}}}$. Again, this implies that $a_{i+1}^{l_{i+1}}\cdots a_{j-1}^{l_{j-1}}a_j=\bfx^{(i)}_{\sparenv{l_1+\cdots+l_i,l_1+\cdots+l_{j-1}}}=\bbz_{\sparenv{l_1+\cdots+l_i,l_1+\cdots+l_{j-1}}}=\bfx^{(j)}_{\sparenv{l_1+\cdots+l_i,l_1+\cdots+l_{j-1}}}=a_{i}a_{i+1}^{l_{i+1}}\cdots a_{j-1}^{l_{j-1}}$, which is a contradiction. This completes the proof of the lower bound.
\end{IEEEproof}

\subsection{Bounds on the size of $s$-deletion-$t$-transposition balls}
It is important to notice that \Cref{thm_DTballsize} and the lower bound established in \Cref{thm_bound1delttrans} rely on \Cref{lem_Hammingdistance}. For general $s\ge 2$, we do not know if an analogous result exists. Therefore, a different idea will be used to bound the size of $\cB_{s,t}(\bfx)$ from below. In simpler terms, we write $\bfx$ as the concatenation of two substrings $\bbu$ and $\bbv$. That is, $\bfx=\bbu\bbv$. Then we apply $s$ deletions on $\bbv$ and $t$ transpositions on $\bbv$.
\begin{lemma}\cite[eq. (11)]{Levenshtein2001JCTA}\cite[Theorem 3.4]{Hirschberg2000JDA}\label{lem_lowerboundsdeletion}
    For any $\bfx\in\Sigma_q^n$, we have
    $$
    \binom{r(\bfx)-s+1}{s}\le\sum_{i=0}^{s}\binom{r(\bfx)-s}{i}\le\abs{\cD_s(\bfx)}\le\binom{r(\bfx)+s-1}{s}.
    $$
\end{lemma}
\begin{theorem}\label{thm_boundsdelttrans}
    Let $s,t\ge1$ be integers and $\bfx\in\Sigma_q^n$ be a sequence with $r$ runs. Then we have
    $$
    \abs{\cB_{s,t}(\bfx)}\le \binom{r(\bfx)+s-1}{s}\prod_{i=0}^{t-1}\parenv{r+2i}.
    $$
    When $r\ge 4t+2$. It holds that
    $$
    \abs{\cB_{s,t}(\bfx)}\ge \sum_{i=0}^{s}\binom{\floorenv{\frac{r}{2}}-s}{i}\cdot\sum_{j=0}^{t}\binom{\ceilenv{\frac{r-2}{4}}}{j}\binom{\ceilenv{\frac{r-2}{4}}-2j-1}{t-j}.
    $$
    In particular, we have $\abs{\cB_{s,t}(\bfx)}\ge\parenv{\frac{r-1-2s}{2s}}^s\parenv{\frac{r-2}{4t}}^t$ when $r\ge \max\mathset{4s+1,4t+2}$.
\end{theorem}
\begin{IEEEproof}
By \Cref{lem_order}, we have $\cB_{s,t}(\bfx)=\cT_{\le t}\parenv{\cD_s(\bfx)}$. Now the upper bound can be proved following the same argument for proving the upper bound in \Cref{thm_bound1delttrans}.

Now we prove the lower bound. Let $l$ be the smallest integer such that $\bfx_{[1,l]}$ has $\floorenv{r/2}$ runs and let $\bbu=\bfx_{[1,l]}$. Let $\bbv=\bfx_{[l+1,n]}$ and $r_1=r\parenv{\bfx^{(2)}}$. It is clear that $\floorenv{r/2}+r_1\in\{r,r+1\}$ and therefore, $r_1\ge\ceilenv{r/2}$. Define
    $$
    \cS=\mathset{\bbu^\prime\bbv^\prime:\bbu^\prime\in\cD_s\parenv{\bbu},\bbv^\prime\in\cT_{t}^\prime\parenv{\bbv}}.
    $$
    Since $\abs{\cS}=\abs{\cD_s\parenv{\bbu}}\cdot\abs{\cT_{t}^\prime\parenv{\bbv}}$, the conclusion follows from \Cref{lem_lowerboundttransposition,lem_lowerboundsdeletion}.
\end{IEEEproof}

\section{Upper bound on the size of $s$-deletion-$t$-transposition codes}\label{sec_bounddeltrans}
In this section, we will use the framework in \Cref{sec_framework} to derive upper bounds on the size of $1$-deletion-$1$-transposition codes.
Before that, we need the next lemma.
\begin{lemma}\label{lem_runchange}
  Let $s,t\ge1$ be integers and $\bfx\in\Sigma_q^{n}$ be a sequence, where $n\ge2$. If $\bby\in\cB_{s,t}(\bfx)$, then $r(\bby)\le r(\bfx)+2t$.
\end{lemma}
\begin{IEEEproof}
    It is easy to see that a deletion does not increase the number of runs and an adjacent transposition can increase the number of runs by at most two. Now the conclusion follows.
\end{IEEEproof}

With above preparation, we are now ready to derive our upper bounds.
\subsection{$1$-deletion-$1$-transposition codes}
For $q\ge2$ and $u\ge4$, define
\begin{equation*}
    \lambda_{q,u}(n)=q\sum_{r=0}^{3}\binom{n-2}{r}(q-1)^{r}+q\sum_{r=4}^{u}\frac{(q-1)^{r}}{(r-1)(r-6)+9}\binom{n-2}{r}-\frac{(u+2)(u+3)q}{\sparenv{u(u-5)+9}n(n-1)(q-1)^2}\sum_{r=0}^{u+2}\binom{n}{r}(q-1)^r.
\end{equation*}
When $q$ and $u$ are fixed, we have $\lambda_{q,u}(n)=\Theta(n^u)$.
\begin{theorem}\label{thm_1del1trans}
For given integers $q\ge 2,u\ge4$ and real number $0<\epsilon<1$, let $n_{q,u,\epsilon}$ be the smallest integer such that $\lambda_{q,u}(n)\le\frac{(u+2)(u+3)q}{\sparenv{u(u-5)+9}(q-1)^2}\cdot\frac{q^n}{n(n-1)}$ for all $n\ge n_{q,u,\epsilon}$. Let $\cC\subseteq\Sigma_q^n$ be a single-deletion-single-transposition code. Then 
    $$
    \abs{\cC}\le \frac{(1+\epsilon)(u+2)(u+3)q}{\sparenv{u(u-5)+9}(q-1)^2}\cdot\frac{q^n}{n(n-1)}
    $$
for all $n\ge n_{q,u,\epsilon}$.
\end{theorem}
\begin{IEEEproof}
    For $\bby\in\Sigma_q^{n-1}$, let 
    \begin{equation*}
        w_{\bby}=
        \begin{cases}
            1,\mbox{ if }r(\bby)\le 4,\\
            \frac{1}{(r(\bby)-2)(r(\bby)-7)+9},\mbox{ if }r(\bby)\ge5.
        \end{cases}
    \end{equation*}
Note that $(r(\bby)-2)(r(\bby)-7)+9>0$ for all $r(\bby)$. Therefore, variables $w_{\bby}$ are well-defined.
    
    Let $\bfx\in\Sigma_q^n$. If there is some $\bby\in\cB_{1,1}(\bfx)$ with $r(\bby)\le 4$, it is clear that $\sum_{\bby\in\cB_{1,1}(\bfx)}w_{\bby}\ge1$. Suppose now that $r(\bby)\ge5$ for all $\bby\in\cB_{1,1}(\bfx)$. By \Cref{lem_runchange}, we have $r(\bby)\le r(\bfx)+2$. Since $r(\bby)\ge 5$, it must be that $r(\bfx)\ge 3$. Combining this with \Cref{thm_DTballsize}, we conclude that
    $$
    \sum_{\bby\in\cB_{1,1}(\bfx)}w_{\bby}=\sum_{\bby\in\cB_{1,1}(\bfx)}\frac{1}{(r(\bby)-2)(r(\bby)-7)+9}\ge\sum_{\bby\in\cB_{1,1}(\bfx)}\frac{1}{r(\bfx)(r(\bfx)-5)+9}=\frac{\abs{\cB_{1,1}(\bfx)}}{r(\bfx)(r(\bfx)-5)+9}\ge1.
    $$
    Now it follows from (\ref{eq_spherepacking2}) that
    \begin{equation*}
    \begin{aligned}
        \abs{\cC}&\le\sum_{\bby\in\Sigma_q^{n-1}}w_{\bby}\\
        &=\sum_{r(\bby)=1}^{4}1+\sum_{r(\bby)=5}^{n-1}\frac{1}{(r(\bby)-2)(r(\bby)-7)+9}\\
        &\overset{(a)}{=}q\sum_{r=1}^{4}\binom{n-2}{r-1}(q-1)^{r-1}+q\sum_{r=5}^{n-1}\frac{(q-1)^{r-1}}{(r-2)(r-7)+9}\binom{n-2}{r-1}\\
        &=q\sum_{r=0}^{3}\binom{n-2}{r}(q-1)^{r}+q\sum_{r=4}^{n-2}\frac{(q-1)^{r}}{(r-1)(r-6)+9}\binom{n-2}{r},
    \end{aligned}
    \end{equation*}
where (a) follows from the fact that there are $\binom{m-1}{r-1}q(q-1)^{r-1}$ $q$-ary length-$m$ sequences with exactly $r$ runs. For any fixed $u\ge 4$, let $\mu(n,q,u)=q\sum_{r=0}^{3}\binom{n-2}{r}(q-1)^{r}+q\sum_{r=4}^{u}\frac{(q-1)^{r}}{(r-1)(r-6)+9}\binom{n-2}{r}$ and $\nu(u)=\frac{(u+2)(u+3)}{u(u-5)+9}$. We obtain
\begin{equation}\label{eq_upperbound1}
    \begin{aligned}
        \abs{\cC}&\le\mu(n,q,u)+q\sum_{r=u+1}^{n-2}\frac{(q-1)^{r}}{(r-1)(r-6)+9}\binom{n-2}{r}\\
        &=\mu(n,q,u)+\frac{q}{n(n-1)}\sum_{r=u+1}^{n-2}\frac{(r+1)(r+2)(q-1)^{r}}{(r-1)(r-6)+9}\binom{n}{r+2}\\
        &\overset{(b)}{\le}\mu(n,q,u)+\frac{\nu(u)q}{n(n-1)}\sum_{r=u+1}^{n-2}\binom{n}{r+2}(q-1)^{r+2}\\
        &=\mu(n,q,u)+\frac{\nu(u)q}{n(n-1)(q-1)^2}\sum_{r=u+3}^{n}\binom{n}{r}(q-1)^{r}\\
        &=\mu(n,q,u)+\frac{\nu(u)q}{n(n-1)(q-1)^2}\sparenv{\sum_{r=0}^{n}\binom{n}{r}(q-1)^{r}-\sum_{r=0}^{u+2}\binom{n}{r}(q-1)^r}\\
        &=\mu(n,q,u)-\frac{\nu(u)q}{n(n-1)(q-1)^2}\sum_{r=0}^{u+2}\binom{n}{r}(q-1)^r+\frac{\nu(u)q}{(q-1)^2}\cdot\frac{q^n}{n(n-1)}.
    \end{aligned}
\end{equation}
Here, inequality (b) follows from the fact that $\frac{(r+1)(r+2)}{(r-1)(r-6)+9}\le \nu(u)$ when $r\ge u+1$.

Since $\mu(n,q,u)-\frac{\nu(u)q}{n(n-1)(q-1)^2}\sum_{r=0}^{u+2}\binom{n}{r}(q-1)^r=\lambda_{q,u}(n)$, the conclusion follows from \Cref{eq_upperbound1} and the definition of $n_{q,t,\epsilon}$.
\end{IEEEproof}

\begin{remark}
    Let $\epsilon$ be fixed. When $u$ increases, the value of $\frac{(u+2)(u+3)q}{\sparenv{u(u-5)+9}(q-1)^2}$ decreases to $\frac{q}{(q-1)^2}$, while the value of $n_{q,\epsilon,u}$ increases.
\end{remark}

\subsection{$1$-deletion-$t$-transposition codes}
Denote $[m]_t\triangleq m(m-1)\cdots(m-t)$ for any $1\le t<m$. For $q,t\ge2$ and $u\ge 10t+2$, define
\begin{align*}
\lambda_{q,t,u}(n)&=q\sum_{r=0}^{10t+1}\binom{n-2}{r}(q-1)^{r}+\frac{q(4t)^t}{[n+t-1]_t}\sum_{r=10t+2}^{u}\frac{[r+t+1]_t(q-1)^r}{(r+1-2t)(r-6t-2)^t}\binom{n+t-1}{r+t+1}\\
&\quad\quad\quad\quad-\frac{q(4t)^t}{[n+t-1]_t(q-1)^{t+1}}\cdot\frac{[u+t+2]_t}{(u+2-2t)(u-6t-1)^t}\sum_{r=0}^{u+t+1}\binom{n+t-1}{r}(q-1)^{r}.
\end{align*}
When $q,t,u$ are fixed, we have $\lambda_{q,t,u}(n)=\Theta\parenv{n^{u}}.$

\begin{theorem}\label{thm_1delmultitrans}
For given integers $q\ge 2$, $t\ge1$, $u\ge 10t+2$ and real number $0<\epsilon<1$, let $n_{q,t,u,\epsilon}$ be the smallest integer such that $\lambda_{q,t,u}(n)\le\epsilon\frac{(4qt)^t[u+t+2]_t}{(q-1)^{t+1}(u+2-2t)(u-6t-1)^t}\cdot\frac{q^{n}}{[n+t-1]_t}$ for all $n\ge n_{q,t,u,\epsilon}$. Let $\cC\subseteq\Sigma_q^n$ be a single-deletion-single-transposition code. Then 
    $$
    \abs{\cC}\le (1+\epsilon)\frac{(4qt)^t[u+t+2]_t}{(q-1)^{t+1}(u+2-2t)(u-6t-1)^t}\cdot\frac{q^{n}}{[n+t-1]_t}
    $$
for all $n\ge n_{q,t,u,\epsilon}$.
\end{theorem}
\begin{IEEEproof}
For $\bby\in\Sigma_q^{n-1}$, let 
    \begin{equation*}
        w_{\bby}=
        \begin{cases}
            1,\mbox{ if }r(\bby)\le 10t+2,\\
            \frac{1}{(r(\bby)-2t)\parenv{\frac{r(\bby)-6t-3}{4t}}^t},\mbox{ if }r(\bby)\ge 10t+3.
        \end{cases}
    \end{equation*}

    Let $\bfx\in\Sigma_q^n$. If there is some $\bby\in\cB_{1,t}(\bfx)$ with $r(\bby)\le 10t+2$, it is clear that $\sum_{\bby\in\cB_{1,t}(\bfx)}w_{\bby}\ge1$. Suppose now that $r(\bby)\ge 10t+3$ for all $\bby\in\cB_{1,t}(\bfx)$. By \Cref{lem_runchange}, we have $r(\bby)\le r(\bfx)+2t$. Since $r(\bby)\ge 10t+3$, it must be that $r(\bfx)\ge 8t+3$. Combining this with \Cref{thm_bound1delttrans}, we conclude that
    $$
    \sum_{\bby\in\cB_{1,t}(\bfx)}w_{\bby}=\sum_{\bby\in\cB_{1,t}(\bfx)}\frac{1}{(r(\bby)-2t)\parenv{\frac{r(\bby)-6t-3}{4t}}^t}\ge\sum_{\bby\in\cB_{1,t}(\bfx)}\frac{1}{r(\bfx)\parenv{\frac{r(\bfx)-4t-3}{4t}}^t}=\frac{\abs{\cB_{1,t}(\bfx)}}{r(\bfx)\parenv{\frac{r(\bfx)-4t-3}{4t}}^t}\ge1.
    $$
    Now it follows from (\ref{eq_spherepacking2}) that
    \begin{equation*}
    \begin{aligned}
        \abs{\cC}&\le\sum_{\bby\in\Sigma_q^{n-1}}w_{\bby}\\
        &=\sum_{r(\bby)=1}^{10t+2}1+\sum_{r(\bby)=10t+3}^{n-1}\frac{1}{(r(\bby)-2t)\parenv{\frac{r(\bby)-6t-3}{4t}}^t}\\
        &=q\sum_{r=1}^{10t+2}\binom{n-2}{r-1}(q-1)^{r-1}+q(4t)^t\sum_{r=10t+3}^{n-1}\frac{(q-1)^{r-1}}{(r-2t)(r-6t-3)^t}\binom{n-2}{r-1}\\
        &=q\sum_{r=0}^{10t+1}\binom{n-2}{r}(q-1)^{r}+q(4t)^t\sum_{r=10t+2}^{n-2}\frac{(q-1)^{r}}{(r+1-2t)(r-6t-2)^t}\binom{n-2}{r}.
    \end{aligned}
    \end{equation*}
   For any fixed $u\ge 10t+2$, denote
    $$
    \mu(n,q,t,u)=q\sum_{r=0}^{10t+1}\binom{n-2}{r}(q-1)^{r}+\frac{q(4t)^t}{[n+t-1]_t}\sum_{r=10t+2}^{u}\frac{(r+1)\cdots(r+t+1)}{(r+1-2t)(r-6t-2)^t}\binom{n+t-1}{r+t+1}(q-1)^{r}
    $$
    and $\nu(t,u)=\frac{[u+t+2]_t}{(u+2-2t)(u-6t-1)^t}$. Then
    \begin{equation}\label{eq_upperbound2}
    \begin{aligned}
       \abs{\cC}&\le \mu(n,q,t,u)+\frac{q(4t)^t\nu(t,u)}{[n+t-1]_t}\sum_{r=u+1}^{n-2}\binom{n+t-1}{r+t+1}(q-1)^{r}\\
        &=\mu(n,q,t,u)+\frac{q(4t)^t\nu(t,u)}{[n+t-1]_t(q-1)^{t+1}}\sum_{r=u+t+2}^{n+t-1}\binom{n+t-1}{r}(q-1)^{r}\\
        &=\mu(n,q,t,u)+\frac{q(4t)^t\nu(t,u)}{[n+t-1]_t(q-1)^{t+1}}\sparenv{\sum_{r=0}^{n+t-1}\binom{n+t-1}{r}(q-1)^{r}-\sum_{r=0}^{u+t+1}\binom{n+t-1}{r}(q-1)^{r}}\\
        &=\mu(n,q,t,u)-\frac{q(4t)^t\nu(t,u)}{[n+t-1]_t(q-1)^{t+1}}\sum_{r=0}^{u+t+1}\binom{n+t-1}{r}(q-1)^{r}+\frac{(4qt)^t\nu(t,u)}{(q-1)^{t+1}}\cdot\frac{q^{n}}{[n+t-1]_t}.
    \end{aligned}
    \end{equation}
    Now the conclusion follows by noticing that $\mu(n,q,t,u)-\frac{q(4t)^t\nu(t,u)}{[n+t-1]_t(q-1)^{t+1}}\sum_{r=0}^{u+t+1}\binom{n+t-1}{r}(q-1)^{r}=\lambda_{q,t,u}(n)$.
\end{IEEEproof}
\begin{remark}
    When $t=1$, the upper bound in \Cref{thm_1delmultitrans} is $(1+\epsilon)\frac{4(u+2)(u+3)q}{u(u-7)(q-1)^2}\cdot\frac{q^n}{n(n-1)}$, which is approximately four times the upper bound in \Cref{thm_1del1trans}.
\end{remark}

\subsection{$s$-deletion-$t$-transposition codes}
For $s,t\ge1$, denote $r_{s,t}\triangleq\max\mathset{4s+2t+1,6t+2}$. Recall that $[m]_t=m(m-1)\cdots(m-t)$. For $q\ge2$, $s,t\ge1$ and $u\ge r_{s,t}-1$, denote $\nu(s,t,u)=\frac{[u+2s+2t+1]_{s+t-1}}{(u-s-t+1)^s(u+s-t)^t}$ and
\begin{align*}
        \lambda_{q,s,t,u}(n)&=q\sum_{r=0}^{r_{s,t}-2}\binom{n-s-1}{r}(q-1)^r+\sum_{r=r_{s,t}-1}^{u}\frac{q(2s)^s(4t)^t(q-1)^r}{(r-2s-2t)^s(r-1-2t)^t}\binom{n-s-1}{r}\\
        &\quad-\frac{q(2s)^s(4t)^t}{(q-1)^{s+t}}\cdot\frac{\nu(s,t,u)}{[n+t-1]_{s+t-1}}\sum_{r=0}^{u+s+t}\binom{n+t-1}{r}(q-1)^r.
    \end{align*}
When $q,s,t$ and $u$ are fixed, we have $\lambda_{q,s,t,u}(n)=\Theta\parenv{n^u}$.
\begin{theorem}\label{thm_multidelmultitran}
For given integers $q\ge 2$, $s,t\ge1$, $u\ge r_{s,t}-1$ and real number $0<\epsilon<1$, let $n_{q,s,t,u,\epsilon}$ be the smallest integer such that $\lambda_{q,s,t,u}(n)\le\epsilon\cdot\frac{(2s)^s(4qt)^t\nu(s,t,u)}{(q-1)^{s+t}}\cdot\frac{q^{n}}{[n+t-1]_{s+t-1}}$ for all $n\ge n_{q,s,t,u,\epsilon}$. Let $\cC\subseteq\Sigma_q^n$ be an $s$-deletion-$t$-transposition code. Then 
    $$
    \abs{\cC}\le (1+\epsilon)\frac{(2s)^s(4qt)^t\nu(s,t,u)}{(q-1)^{s+t}}\cdot\frac{q^{n}}{[n+t-1]_{s+t-1}}
    $$
for all $n\ge n_{q,s,t,u,\epsilon}$.
\end{theorem}
\begin{IEEEproof}
For $\bby\in\Sigma_q^{n-s}$, let 
    \begin{equation*}
        w_{\bby}=
        \begin{cases}
            1,\mbox{ if }r(\bby)<r_{s,t},\\
            \frac{1}{\parenv{\frac{r(\bby)-2s-2t-1}{2s}}^s\parenv{\frac{r(\bby)-2t-2}{4t}}^t},\mbox{ if }r(\bby)\ge r_{s,t}.
        \end{cases}
    \end{equation*}

    Let $\bfx\in\Sigma_q^n$. If there is some $\bby\in\cB_{s,t}(\bfx)$ with $r(\bby)< r_{s,t}$, it is clear that $\sum_{\bby\in\cB_{1,t}(\bfx)}w_{\bby}\ge1$. Suppose now that $r(\bby)\ge r_{s,t}$ for all $\bby\in\cB_{s,t}(\bfx)$. By \Cref{lem_runchange}, we have $r(\bby)\le r(\bfx)+2t$. Since $r(\bby)\ge r_{s,t}$, it must be that $r(\bfx)\ge\max\mathset{4s+1,4t+2}$. Combining this with \Cref{thm_boundsdelttrans}, we conclude that
    \begin{align*}
        \sum_{\bby\in\cB_{s,t}(\bfx)}w_{\bby}&=\sum_{\bby\in\cB_{s,t}(\bfx)}\frac{1}{\parenv{\frac{r(\bby)-2s-2t-1}{2s}}^s\parenv{\frac{r(\bby)-2t-2}{4t}}^t}\\
        &\ge\sum_{\bby\in\cB_{s,t}(\bfx)}\frac{1}{\parenv{\frac{r(\bfx)-2s-1}{2s}}^s\parenv{\frac{r(\bfx)--2}{4t}}^t}\\
        &=\frac{\abs{\cB_{s,t}(\bfx)}}{\parenv{\frac{r(\bby)-2s-2t-1}{2s}}^s\parenv{\frac{r(\bby)-2t-2}{4t}}^t}\ge1.
    \end{align*}
    
    Then by (\ref{eq_spherepacking2}) and following similar arguments in proofs of \Cref{thm_1del1trans,thm_1delmultitrans}, we obtain
    \begin{equation}\label{eq_upperbound3}
        \begin{aligned}
            \abs{\cC}&\le\lambda_{q,s,t,u}(n)+\frac{(2s)^s(4qt)^t\nu(s,t,u)}{(q-1)^{s+t}}\cdot\frac{q^{n}}{[n+t-1]_{s+t-1}}.
        \end{aligned}
    \end{equation}
    Now the conclusion follows from (\ref{eq_upperbound3}) and the definition of $n_{q,s,t,u,\epsilon}$.
\end{IEEEproof}
\begin{remark}
    Let $s=1$. The upper bound in \Cref{thm_multidelmultitran} is roughly two times the upper bound in \Cref{thm_1delmultitrans}.
\end{remark}

\section{Upper Bound on Codes Correcting Block-Deletions and adjacent Block-Transpositions}\label{sec_boundblock}
We generalize notions of deletions and adjacent transpositions to their block-level counterparts. Let $s$, $t$, $b$ and $n$ be three positive integers. Let $\bfx\in\Sigma_q^n$ and $\bby\in\Sigma_q^{n-sb}$, where $n>sb$. We say that $\bby$ is obtained from $\bfx$ by $s$ $b$-\emph{block deletions}, if $\bby=\bfx_{[n]\setminus\cup_{i=1}^sI_i}$, where $I_1,\ldots,I_s$ are $s$ mutually disjoint intervals of length $b$ of $[n]$. That is to say, $\bby$ is obtained from $\bfx$ by deleting $s$ non-overlapping substrings of length $b$. Let $\bfx, \bbz\in\Sigma_q^n$, where $n\ge 2b$. We say that $\bbz$ is obtained from $\bfx$ by one $b$-\emph{adjacent block transposition} (or $b$-\emph{block transposition}, for short), if $\bbz=\bfx_{[1,i-1]}\bfx_{[i+b,i+2b-1]}\bfx_{[i,i+b-1]}\bfx_{[i+2b,n]}$. In other words, the two adjacent substrings $\bfx_{[i,i+b-1]}$ and $\bfx_{[i+b,i+2b-1]}$ are swapped.

For integers $s,t,b\ge1$, $n\ge(s+2)b$ and a sequence $\bfx\in\Sigma_q^n$, define
$$
\cB_{s,t}^{b}(\bfx)=\mathset{\bby\in\Sigma_q^{n-sb}:
\begin{array}{c}
     \bby\text{ is obtained from }\bfx\text{ by }s\\
     b\text{-block deletions and at most}\\
     t\text{ }b\text{-block transpositions}
\end{array}
}.
$$
\begin{definition}\label{dfn_blockcode}
  Let $\mathcal{C}\subseteq\Sigma_q^n$. If $\cB_{s,t}^{b}(\bfx)\cap \cB_{s,t}^{b}(\bby)=\emptyset$ for any two distinct sequences $\bfx$ and $\bby$ in $\mathcal{C}$, we call $\mathcal{C}$ an $(s,t,b)$-block-deletion-transposition correcting code.
\end{definition}

Applying \Cref{thm_boundsdelttrans} and following similar idea in the proof of \cite[Theorem III.2]{Zuoye202408arXiv}, we can obtain the following theorem. Briefly speaking, a code $\cC$ is partitioned into two parts $\cC_1$ and $\cC_2$, such that $\abs{\cB_{s,t}^{b}(\bfx)}$ is sufficiently large for each $\bfx\in\cC_1$ and $\abs{\cC_2}$ is sufficiently small. Then a packing argument can be applied to $\cC_1$ to get an upper bound on $\abs{\cC_1}$.
\begin{theorem}\label{thm_upperboundblock}
 Let $q\ge2,s,t,b\ge1$ and $n\ge (s+2)b$ be integers. Let $\cC\in\Sigma_q^n$ be an $(s,t,b)$-block-deletion-transposition correcting code. For $q\ge2$, let $f(q)=\min\mathset{\frac{1}{q},\frac{q-1}{2q},\frac{(q-1)^2}{q^2-3q+6}\parenv{\frac{1}{q}-\frac{(q-1)\ln q}{2q^3}}}$. Let $0<\mu<1$ be a real number. Suppose that $n$ is sufficiently large such that $\parenv{1-\frac{\epsilon q}{q-1}}^{s+t}\parenv{1-\frac{b}{n}}^{s+t}\ge\mu$. Then it holds that
    \begin{equation*}
        \abs{\cC}\le\parenv{\frac{(2s)^s(4t)^t(bq)^{s+t}}{\mu q^{sb}(q-1)^{s+t}}+\frac{(1.21)^{(s+t+1)b}}{n}}\frac{q^n}{n^{s+t}}.
    \end{equation*}
\end{theorem}
\begin{IEEEproof}
    If $b\nmid n$, we can define a set
    $$
    \cC^\prime=\mathset{\bfx_{\sparenv{1,\floorenv{n/b}b}}:\bfx\in\cC}.
    $$
    Since $\cC$ can correct $s$ $b$-block deletions, we have $\abs{\cC^\prime}=\abs{\cC}$. Moreover, since $\cC$ is an $(s,t,b)$-block-deletion-transposition correcting code, $\cC^\prime$ is also an $(s,t,b)$-block-deletion-transposition correcting code. Therefore, we can always assume that $b\mid n$. In this case, we represent each codeword $\bfx\in\cC$ as a $b\times n/b$ array as follows:
    \begin{equation*}
    A(\bfx)=
        \begin{pmatrix}
            x_1&x_{b+1}&\cdots & x_{n-b+1}\\
            x_2&x_{b+2}&\cdots & x_{n-b+2}\\
            \vdots&\vdots&\cdots&\vdots\\
            x_b&x_{2b}&\cdots& x_n
        \end{pmatrix}.
    \end{equation*}
   For each $i$, let $A(\bfx)_i$ be the $i$-th row of $A(\bfx)$. For $\bfx\in\cC$, let
    $$
    \cA(\bfx)=\mathset{A^{\prime}(\bfx):
    \begin{array}{c}
       A^{\prime}(\bfx)\text{ is obtained from }A(\bfx)\text{ by}\\
    \text{deleting }s\text{ columns and transposing}\\
    \text{at most }t\text{ adjacent columns}
    \end{array}
    }.
    $$
    It is easy to verify that $\cup_{A^\prime\in\cA(\bfx)}\mathset{A^\prime_i}=\cB_{s,t}\parenv{A(\bfx)_i}$ for each $i$. Then it follows that
    \begin{equation}\label{eq_blockupperbound1}
    \begin{aligned}
            \abs{\cB_{s,t}^{b}(\bfx)}\ge\abs{\cA(\bfx)}&\ge\max_{1\le i\le b}\mathset{\abs{\cup_{A^\prime\in\cA(\bfx)}\mathset{A^\prime_i}}}\\
            &=\max_{1\le i\le b}\mathset{\abs{\cB_{s,t}\parenv{A(\bfx)_i}}}\\
            &\overset{(a)}{\ge}\max_{1\le i\le b}\mathset{\binom{r\parenv{A(\bfx)_i}-1-2s}{2s}^s\binom{r\parenv{A(\bfx)_i}-2}{4t}^t}\\
            &\ge\max_{1\le i\le b}\mathset{\parenv{\frac{r\parenv{A(\bfx)_i}-1-2s}{2s}}^s\parenv{\frac{r\parenv{A(\bfx)_i}-2}{4t}}^t},
    \end{aligned}
    \end{equation}
    where the (a) follows from \Cref{thm_boundsdelttrans}.

    Set $m=n/b-1$, $\epsilon=\sqrt{\frac{4(s+t+1)\log n}{n\log q}}$ and $r_0=\parenv{1-\frac{1}{q}-\epsilon}m+2s$. Partition $\cC$ into two parts $\cC=\cC_1\cup\cC_2$, where $\cC_1=\mathset{\bfx\in\cC:r\parenv{A(\bfx)_i}>r_0\text{ for some }i}$ and $\cC_2=\mathset{\bfx\in\cC:r\parenv{A(\bfx)_i}\le r_0\text{ for all }i}$. Then we have $\abs{\cC}=\abs{\cC_1}+\abs{\cC_2}$. The aim is to upper bound $\abs{\cC_1}$ and $\abs{\cC_2}$.

   Since $\cC$ is an $(s,t,b)$-block-deletion-transposition code, $\cC_1$ is also an $(s,t,b)$-block-deletion-transposition code. It follows from \Cref{dfn_blockcode} and (\ref{eq_blockupperbound1}) that
   $$
   \abs{\cC_1}\parenv{\frac{\parenv{1-\frac{1}{q}-\epsilon}m}{2s}}^s\parenv{\frac{\parenv{1-\frac{1}{q}-\epsilon}m}{4t}}^t \le\sum_{\bfx\in\cC_1}\cB_{s,t}^{b}(\bfx)\le q^{n-sb}.
   $$
   Therefore, we have
   \begin{align*}
       \abs{\cC_1}&\le\frac{(2s)^s(4t)^tq^{n-sb}}{\parenv{1-\frac{1}{q}-\epsilon}^{s+t}\parenv{\frac{n}{b}-1}^{s+t}}\\
       &=\frac{(2s)^s(4t)^tq^{n-sb}}{n^{s+t}}\cdot\frac{n^{s+t}}{\parenv{1-\frac{1}{q}-\epsilon}^{s+t}\parenv{\frac{n}{b}-1}^{s+t}}\\
       &=\frac{(2s)^s(4t)^tq^{n-sb}}{n^{s+t}}\cdot\parenv{\frac{bq}{q-1}}^{s+t}\cdot\frac{1}{\parenv{1-\frac{\epsilon q}{q-1}}^{s+t}\parenv{1-\frac{b}{n}}^{s+t}}\\
       &\overset{(b)}{\le}\frac{(2s)^s(4t)^t(bq)^{s+t}}{\mu q^{sb}(q-1)^{s+t}}\cdot\frac{q^n}{n^{s+t}},
   \end{align*}
   where (b) follows from the fact that $\parenv{1-\frac{\epsilon q}{q-1}}^{s+t}\parenv{1-\frac{b}{n}}^{s+t}\ge\mu$.

   We have proved an upper bound for $\abs{\cC_1}$. Next, we upper bound $\abs{\cC_2}$. By the choice of $r_0$, we have $r_0-1\le(1-1/q-\epsilon)m$. Then following similar argument in the proof of  \cite[Theorem III.2]{Zuoye202408arXiv}, we can show that $\abs{\cC_2}\le \frac{(1.21)^{(s+t+1)b}q^n}{n^{s+t+1}}$. Now the proof is completed.
\end{IEEEproof}

\begin{remark}
    A code $\cC$ is called an $(s,t,\le b)$-block-deletion-transposition correcting code if for any $b^\prime\le b$ it is an $(s,t,b^\prime)$-block-deletion-transposition correcting code. Ryan \textit{et al} \cite{Ryan2018IT} constructed a  $(1,1,\le b)$-block-deletion-transposition correcting code with redundancy $\ceilenv{\log b}\log n+O\parenv{b^2\log\log n}$. On the other hand, the lower bound on redundancy implied by \Cref{thm_upperboundblock} is $2\log n-O(1)$.
\end{remark}

\section{Extension to Codes Tolerating Insertions and Substitutions}\label{sec_boundextension}
Let $\bfx=x_1\cdots x_n\in\Sigma_q^n$. A \emph{substitution} at position $i$ means the replacement of $x_i$ with a symbol in $\Sigma_q\setminus\{x_i\}$. An \emph{insertion} at position $i$ refers to inserting a symbol $a\in\Sigma_q$ between $x_{i-1}$ and $x_{i}$. When $i=0$, it means inserting $a$ on the left of $x_1$ and when $i=n+1$, it means inserting $a$ on the right of $x_{n}$. In this section, we briefly explain how to extend upper bounds in \Cref{sec_bounddeltrans} to the case where in addition to deletions and transpositions, insertions and substitutions also occur.

For non-negative integers $s_{\rm{D}}$, $s_{\rm{I}}$, $t_{\rm{T}}$ and $t_{\rm{S}}$, denote by $\cB_{s_{\rm{D}},s_{\rm{I}},t_{\rm{T}},t_{\rm{S}}}^{(q)}(\bfx)$ the set of all sequences that can be obtained from $\bfx$ by exactly $s_{\rm{D}}$ deletions, exactly $s_{\rm{I}}$ insertions, at most $t_{\rm{T}}$ transpositions and at most $t_{\rm{S}}$ substitutions.
A non-empty set $\cC\subseteq\Sigma_q^n$ is called an $(s_{\rm{D}},s_{\rm{I}},t_{\rm{T}},t_{\rm{S}})$-correcting code if $\cB_{s_{\rm{D}},s_{\rm{I}},t_{\rm{T}},t_{\rm{S}}}^{(q)}(\bfx)\cap\cB_{s_{\rm{D}},s_{\rm{I}},t_{\rm{T}},t_{\rm{S}}}^{(q)}(\bby)=\emptyset$ for any two distinct $\bfx,\bby\in\cC$.

It is well-known that \cite[eq. (24)]{Levenshtein2001JCTA}
\begin{equation}\label{eq_ballsizeinser}
    \abs{\cB_{0,s_{\rm{I}},0,0}^{(q)}(\bfx)}=\sum_{i=0}^{s_{\rm{I}}}\binom{n+s_{\rm{I}}}{i}(q-1)^i
\end{equation}
and
\begin{equation}\label{eq_ballsizesub}
    \abs{\cB_{0,0,0,t_{\rm{S}}}^{(q)}(\bfx)}=\sum_{i=0}^{t_{\rm{S}}}\binom{n}{i}(q-1)^i.
\end{equation}
for any $\bfx\in\Sigma_q^n$.

The idea for proving \Cref{thm_boundsdelttrans} can be generalized to give a lower bound of $\abs{\cB_{s_{\rm{D}},s_{\rm{I}},t_{\rm{T}},t_{\rm{S}}}(\bfx)}$. 
\begin{lemma}\label{lem_extensionlowerbound}
    Let $s_{\rm{D}},s_{\rm{I}},t_{\rm{T}}$ and $t_{\rm{S}}$ be non-negative integers satisfying $s_{\rm{D}}+s_{\rm{I}}+t_{\rm{T}}+t_{\rm{S}}\ge1$. Let $\bfx\in\Sigma_q^n$ be a sequence with $r$ runs. If $r\ge\max\mathset{8s_{\rm{D}}+3,8t_{\rm{T}}+7}$ and $n\ge\max\mathset{8s_{\rm{I}}+3,4t_{\rm{S}}+3}$, we have
    $$
    \abs{\cB_{s_{\rm{D}},s_{\rm{I}},t_{\rm{T}},t_{\rm{S}}}(\bfx)}\ge(q-1)^{s_{\rm{I}}+t_{\rm{S}}}\parenv{\frac{n-3-4s_{\rm{I}}}{4s_{\rm{I}}}}^{s_{\rm{I}}}\parenv{\frac{n-3}{4t_{\rm{S}}}}^{t_{\rm{S}}}\parenv{\frac{r-3-4s_{\rm{D}}}{4s_{\rm{D}}}}^{s_{\rm{D}}}\parenv{\frac{r-7}{8t_{\rm{T}}}}^{t_{\rm{T}}}.
    $$
\end{lemma}
\begin{remark}
    In \Cref{lem_extensionlowerbound,lem_asyballsize} and \Cref{thm_extenedbound,thm_asybound}, we define $\infty^0=1$. In this way, parameters $s_{\rm{D}}$, $s_{\rm{I}}$, $t_{\rm{S}}$, $t_{\rm{T}}$, $s$, $t^{+}$ and $t^{-}$ are allowed to be $0$. 
\end{remark}
\newenvironment{pf1}{{\it Proof of \Cref{lem_extensionlowerbound}:}}{\hfill $\blacksquare$\par}
\begin{pf1}
    Write $\bfx$ as the concatenation of two substrings $\bfx=\bbu\bbv$. It must be that one of two substrings has length at least $\floorenv{n/2}$ and the other substring has at least $\floorenv{r/2}$ runs. Without loss of generality, assume that $\bbu$ has length at least $\floorenv{n/2}$ and $r(\bbv)\ge\floorenv{r/2}$. Next write $\bbu$ as the concatenation of two substrings $\bbu=\bfx_1 \bfx_2$ such that each substring has length at least $\floorenv{n/4}$. Similar to the proof of \Cref{thm_boundsdelttrans}, we can partition $\bbv$ into two parts $\bbv=\bfx_3 \bfx_4$ such that $r\parenv{\bfx_3},r\parenv{\bfx_4}\ge\floorenv{r/4}$. 
    
    Recall that $\cT_{t}^{\prime}(\bfx)$ denotes the set of all sequences obtained from $\bfx$ by exactly $t$ \emph{simultaneous} transpositions. Define
    $$
    \cS=\mathset{\bby=\bby_1\bby_2\bby_3\bby_4:
    \begin{array}{c}
    \bby_1\in\cB_{0,s_{\rm{I}},0,0}\parenv{\bfx_1},\\
    \bby_2\in\cB_{0,0,0,t_{\rm{S}}}\parenv{\bfx_2},\\
    \bby_3\in\cB_{s_{\rm{D}},0,0,0}\parenv{\bfx_3},\\
    \bby_4\in\cT_{t_{\rm{T}}}^\prime\parenv{\bfx_4}.
    \end{array}
    }.
    $$
    Then $\cS\subseteq \cB_{s_{\rm{D}},s_{\rm{I}},t_{\rm{T}},t_{\rm{S}}}(\bfx)$. It is clear that
    \begin{align*}
        \abs{\cS}&=\abs{\cB_{0,s_{\rm{I}},0,0}\parenv{\bfx_1}}\cdot\abs{\cB_{0,0,0,t_{\rm{S}}}\parenv{\bfx_2}}\cdot\abs{\cB_{s_{\rm{D}},0,0,0}\parenv{\bfx_3}}\cdot\abs{\cB_{0,0,t_{\rm{T}},0}\parenv{\bfx_4}}\\
        &\ge\parenv{\sum_{i=0}^{s_{\rm{I}}}\binom{\floorenv{\frac{n}{4}}+s_{\rm{I}}}{i}(q-1)^i}\cdot\parenv{\sum_{i=0}^{t_{\rm{S}}}\binom{\floorenv{\frac{n}{4}}}{i}(q-1)^i}\\
        &\quad\cdot\parenv{\sum_{i=0}^{s_{\rm{D}}}\binom{\floorenv{r/4}-s_{\rm{D}}}{i}}\cdot\parenv{\frac{\floorenv{r/4}-1}{2t_{\rm{T}}}}^{t_{\rm{T}}},
    \end{align*}
    where the inequality follows from \Cref{lem_lowerboundttransposition,lem_lowerboundsdeletion} and \Cref{eq_ballsizeinser,eq_ballsizesub}.
\end{pf1}

Based on \Cref{lem_extensionlowerbound}, we can apply similar idea in the proof of \Cref{thm_multidelmultitran} to prove an upper bound on the size of $(s_{\rm{D}},s_{\rm{I}},t_{\rm{T}},t_{\rm{S}})$-correcting codes. \Cref{thm_multidelmultitran} says that for given $q,s$ and $t$, when $n$ is sufficiently large, the size of an $s$-deletion-$t$-transposition code is upper bounded by $C\frac{q^n}{n^{s+t}}$, for some number $C$ depending only on $q,s$ and $t$.
\begin{theorem}\label{thm_extenedbound}
For fixed non-negative integers $s_{\rm{D}},s_{\rm{I}},t_{\rm{T}}$ and $t_{\rm{S}}$ satisfying $t\triangleq s_{\rm{D}}+s_{\rm{I}}+t_{\rm{T}}+t_{\rm{S}}\ge1$,  let $r_0=\max\mathset{8s_{\rm{D}}+3,8t_{\rm{T}}+7}$. Then there exists an integer $n_0$, such that whenever $n\ge n_0$, there is a number $C$ depending only on $q$, $s_{\rm{D}},s_{\rm{I}},t_{\rm{T}}$ and $t_{\rm{S}}$, such that $\abs{\cC}\le C\frac{q^n}{n^{t}}$ for any $(s_{\rm{D}},s_{\rm{I}},t_{\rm{T}},t_{\rm{S}})$-correcting code $\cC$.
\end{theorem}
\begin{IEEEproof}
    Let $n_{q,s_{\rm{I}},t_{\rm{S}}}=(q-1)^{s_{\rm{I}}+t_{\rm{S}}}\parenv{\frac{n-3-4s_{\rm{I}}}{4s_{\rm{I}}}}^{s_{\rm{I}}}\parenv{\frac{n-3}{4t_{\rm{S}}}}^{t_{\rm{S}}}$. For $\bby\in\Sigma_{q}^{n-s_{\rm{D}}+s_{\rm{I}}}$, denote $r^{*}(\bby)=r(\bby)-2s_{\rm{I}}-2t_{\rm{S}}-2t_{\rm{T}}$ and let
    $$
    w_{\bby}=
    \begin{cases}
        1,\mbox{ if }r(\bby)<r_0+2s_{\rm{I}}+2t_{\rm{S}}+2t_{\rm{T}},\\
        \frac{1}{n_{q,s_{\rm{I}},t_{\rm{S}}}\parenv{\frac{r^{*}(\bby)-3-4s_{\rm{D}}}{4s_{\rm{D}}}}^{s_{\rm{D}}}\parenv{\frac{r^{*}(\bby)-7}{8t_{\rm{T}}}}^{t_{\rm{T}}}},\mbox{ otherwiese}.
    \end{cases}
    $$
    Let $\bfx\in\Sigma_q^n$. If there is some $\bby\in\cB_{s_{\rm{D}},s_{\rm{I}},t_{\rm{T}},t_{\rm{S}}}(\bfx)$ with $r(\bby)<r_0+2s_{\rm{I}}+2t_{\rm{S}}+2t_{\rm{T}}$, then it is clear that $\sum_{\bby\in\cB_{s_{\rm{D}},s_{\rm{I}},t_{\rm{T}},t_{\rm{S}}}(\bfx)}w_{\bby}\ge1$. Now suppose that $r(\bby)\ge r_0+2s_{\rm{I}}+2t_{\rm{S}}+2t_{\rm{T}}$ for all $\bby\in\cB_{s_{\rm{D}},s_{\rm{I}},t_{\rm{T}},t_{\rm{S}}}(\bfx)$. Since one deletion does not increase the number of runs while one insertion (substitution or transposition) increases the number of runs by at most $2$, it holds that $r(\bfx)\ge r^{*}(\bby)\ge r_0$. Then by \Cref{lem_extensionlowerbound}, we have
    \begin{align*}
    \sum_{\bby\in\cB_{s_{\rm{D}},s_{\rm{I}},t_{\rm{T}},t_{\rm{S}}}(\bfx)}w_{\bby}&=\sum_{\bby\in\cB_{s_{\rm{D}},s_{\rm{I}},t_{\rm{T}},t_{\rm{S}}}(\bfx)}\frac{1}{n_{q,s_{\rm{I}},t_{\rm{S}}}\parenv{\frac{r^{*}(\bby)-3-4s_{\rm{D}}}{4s_{\rm{D}}}}^{s_{\rm{D}}}\parenv{\frac{r^{*}(\bby)-7}{8t_{\rm{T}}}}^{t_{\rm{T}}}}\\
    &\ge\sum_{\bby\in\cB_{s_{\rm{D}},s_{\rm{I}},t_{\rm{T}},t_{\rm{S}}}(\bfx)}\frac{1}{n_{q,s_{\rm{I}},t_{\rm{S}}}\parenv{\frac{r(\bfx)-3-4s_{\rm{D}}}{4s_{\rm{D}}}}^{s_{\rm{D}}}\parenv{\frac{r(\bfx)-7}{8t_{\rm{T}}}}^{t_{\rm{T}}}}\\
    &\ge \sum_{\bby\in\cB_{s_{\rm{D}},s_{\rm{I}},t_{\rm{T}},t_{\rm{S}}}(\bfx)}\frac{1}{\abs{\cB_{s_{\rm{D}},s_{\rm{I}},t_{\rm{T}},t_{\rm{S}}}(\bfx)}}=1.
    \end{align*}

    Now the conclusion follows from (\ref{eq_spherepacking2}) and similar argument in the proof of \Cref{thm_1del1trans,thm_1delmultitrans,thm_multidelmultitran}.
\end{IEEEproof}

\begin{remark}
    As in \Cref{thm_1del1trans,thm_1delmultitrans,thm_multidelmultitran}, the values of $n_0$ and $C$ can be specified. We do not do so here for conciseness.
\end{remark}

\section{Upper Bounds on Codes Correcting deletions and Asymmetric Transpositions}\label{sec_boundasy}
In this section, we focus on the binary alphabet $\{0,1\}$. Then there are two types of adjacent transpositions: $0$-right shifts (i.e., $01\rightarrow 10$) and $0$-left shifts (i.e., $10\rightarrow 01$). In previous sections, these two types of transpositions are not distinguished. However, in some scenarios, these two types of transpositions may exhibit non-identical probabilistic characteristics \cite{Nunnelley1990}. In \cite{Shuche2025TCOMM}, Wang \emph{et al} studied the interaction between deletions and asymmetric transposition. They construct a code with $(1+t^{+}+t^{-})\log(1+t^{+}+t^{-})+1$ bits of redundancy, which can correct one deletion, at most $t^{+}$ $0$-right shifts and at most $t^{-}$ $0$-left shifts. In this section, we show that this redundancy is optimal up to a constant.

For integers $s,t^{+},t^{-}\ge 0$ and a sequence $\bfx\in\{0,1\}^n$, let $\cB_{s}^{t^{+},t^{-}}(\bfx)$ denote the set of all sequences which is obtained from $\bfx$ by exactly $s$ deletions, at most $t^{+}$ $0$-right shifts and at most $t^{-}$ $0$-left shifts. As a routine step, we give a lower bound on $\abs{\cB_{s}^{t^{+},t^{-}}(\bfx)}$.
\begin{lemma}\label{lem_asyballsize}
 If $r=r(\bfx)\ge\max\mathset{4s+1,8t^{+}+2}$, it holds that
 \begin{equation*}
 \begin{aligned}
     \abs{\cB_{s}^{t^{+},t^{-}}(\bfx)}&\ge\sum_{i=0}^{s}\binom{r_1-s}{i}\binom{\floorenv{\ceilenv{r/2}/2}}{t^{+}}\binom{\floorenv{\ceilenv{r/2}/2}-2t^{+}}{t^{-}}\\
     &\ge \parenv{\frac{r-1-2s}{2s}}^s\parenv{\frac{r-2}{4t^{+}}}^{t^{+}}\parenv{\frac{r-2-8t^{+}}{4t^{-}}}^{t^{-}}.
 \end{aligned}
 \end{equation*}
\end{lemma}
\begin{IEEEproof}
    As in the proof of \Cref{thm_boundsdelttrans}, we can partition $\bfx$ into two non-overlapping substrings $\bfx=\bbu\bbv$ such that $r_1=r(\bbu)\ge\floorenv{r/2}$ and $r_2=r(\bbv)\ge\ceilenv{r/2}$. Let
    $$
    \cS=\mathset{\bbu^\prime\bbv^\prime:\bbu^\prime\in\cD_s\parenv{\bbu},\bbv^\prime\in\cB_{0}^{t^{+},t^{-}}(\bbv)}.
    $$
    Then we have $\cS\subseteq\cB_{s}^{t^{+},t^{-}}(\bfx)$ and thus,
    \begin{equation*}
    \abs{\cB_{s}^{t^{+},t^{-}}(\bfx)}\ge\abs{\cS}=\abs{\cD_s(\bbu)}\cdot\abs{\cB_{0}^{t^{+},t^{-}}(\bbv)}\overset{(a)}{\ge}\sum_{i=0}^{s}\binom{r_1-s}{i}\cdot\abs{\cB_{0}^{t^{+},t^{-}}(\bbv)},
    \end{equation*}
where (a) follows from \Cref{lem_lowerboundsdeletion}. It remains to show that $\abs{\cB_{0}^{t^{+},t^{-}}(\bbv)}\ge\binom{(r-2)/4}{t^{+}}\binom{(r-2)/4-2t^{+}}{t^{-}}$.
    
    Write $\bbv$ as $\bbv=a_1^{l_1}\cdots a_{r_2}^{l_{r_2}}$, where $a_1^{l_1}\ldots a_{r_2}^{l_{r_2}}$ are all runs in $\bbv$. Denote $R_0=\mathset{i: a_i=0}$. In other words, $R_0$ is the set of indices of all runs of $0$s. Clearly, we have $\abs{R_0}\ge\floorenv{r_2/2}\ge\floorenv{\ceilenv{r/2}/2}\ge(r-2)/4$.

    We choose $t^{+}$ $0$-right shifts and $t^{-}$ $0$-left shifts in the following way. Firstly, choose $i_1,\ldots,i_{t^{+}}\in R_0$. For each $k$, transpose $a_{i_k}$ and $a_{i_k+1}$. This contributes $t^{+}$ $0$-right shifts. Next, choose $j_1,\ldots,j_{t^{-}}\in R_0\setminus\parenv{\mathset{i_1,\ldots,i_{t^{+}}}\cap\mathset{i_1+2,\ldots,i_{t^{+}}+2}}$. Then for each $k$, transpose $a_{j_k}$ and $a_{j_k-1}$. This contributes $t^{-}$ $0$-left shifts.
    
    By the choice of $i_1,\ldots,i_{t^{+}}$ and $j_1,\ldots,j_{t^{-}}$, the $t^{+}+t^{-}$ transposed pairs are mutually non-overlapping. According to \Cref{lem_transdistinct}, different choices of $\parenv{i_1,\ldots,i_{t^{+}},j_1,\ldots,j_{t^{-}}}$ result in different sequences in $\cB_{0}^{t^{+},t^{-}}(\bbv)\cap\cT_{t^{+}+t^{-}}^\prime(\bbv)$. Now the aforementioned lower bound of $\abs{\cB_{0}^{t^{+},t^{-}}(\bbv)}$ follows by noticing that $\abs{\mathset{i_1,\ldots,i_{t^{+}}}\cap\mathset{i_1+2,\ldots,i_{t^{+}}+2}}\le 2t^{+}$.
\end{IEEEproof}

\begin{theorem}\label{thm_asybound}
For non-negative integers $s$, $t^{+}$ and $t^{-1}$ satisfying $t\triangleq s+t^{+}+t^{-}\ge 1$, let $r_0=\max\mathset{4s+1,8t^{+}+2}$. There exists an integer $n_0$, such that whenever $n\ge n_0$, there is a number $C$ depending only on $q$, $s,t^{+}$ and $t^{-}$, such that $\abs{\cC}\le C\frac{q^n}{n^{t}}$ for any code $\cC\subseteq\Sigma_q^n$ capable of correcting $s$ deletions, at most $t^{+}$ $0$-right shifts and at most $t^{-}$ $0$-left shifts.  
\end{theorem}
\begin{IEEEproof}
 For $\bby\in\Sigma_{q}^{n-s}$, denote $r^{*}(\bby)=r(\bby)-2t^{+}-2t^{-}$ and let
    $$
    w_{\bby}=
    \begin{cases}
        1,\mbox{ if }r(\bby)<r_0+2t^{+}+2t^{-},\\
        \frac{1}{\parenv{\frac{r^{*}(\bby)-1-2s}{2s}}^s\parenv{\frac{r^{*}(\bby)-2}{4t^{+}}}^{t^{+}}\parenv{\frac{r^{*}(\bby)-2-8t^{+}}{4t^{-}}}^{t^{-}}},\mbox{ otherwiese}.
    \end{cases}
    $$
By \Cref{lem_asyballsize},  we can verify that $\sum_{\bby\in\cB_{s_{\rm{D}},s_{\rm{I}},t_{\rm{T}},t_{\rm{S}}}(\bfx)}w_{\bby}\ge1$ for all $\bfx\in\Sigma_q^n$. 
Then the conclusion follows from (\ref{eq_spherepacking2}) and similar argument in the proof of \Cref{thm_1del1trans,thm_1delmultitrans,thm_multidelmultitran}.    
\end{IEEEproof}

\begin{remark}
    The above theorem says that when $s$, $t^{+}$ and $t^{-}$ are constants compared to $n$, any code capable of handling $s$ deletions, at most $t^{+}$ $0$-right shifts and at most $t^{-}$ $0$-left shifts requires at least $\parenv{s+t^{+}+t^{-}}\log n-O(1)$ bits of redundancy. As a corollary, the redundancy of the code given in \cite{Shuche2025TCOMM} is optimal up to a constant.
\end{remark}

\section{Conclusion}\label{sec_conclusion}
In this paper, we upper bound the size of codes under the Damerau-Levenshtein metric. Our results show that when the total number $t$ of all errors is a constant, then the redundancy of a code is at least $t\log n-O(1)$. This proves that the redundancy of the code correcting one deletion and asymmetric transpositions constructed in \cite{Shuche2025TCOMM} is optimal up to a constant.

Regarding transpositions and deletions of arbitrary symbols, existing works \cite{Ryan2018IT,Shuche2025TCOMM} focused exclusively on codes correcting one deletion and multiple transpositions. The problem to construct codes capable of correcting multiple deletions and multiple transpositions with redundancy close to $t\log n-O(1)$ is still open. In addition, both works focused on binary codes. It is also interesting to construct codes over non-binary alphabets. At last, constructing codes capable of correcting all four types of errors is also an interesting problem. We leave these three problems for future research.

\bibliographystyle{IEEEtran}
\bibliography{ref}
\end{document}